\documentclass[12pt,preprint]{aastex63}
\usepackage{amssymb}
\usepackage{amsmath}
\usepackage{romannum}
\usepackage{epstopdf}
\usepackage{xcolor}
\usepackage{amsmath}
\usepackage{graphicx}
\usepackage{hyperref}
\usepackage{graphics,graphicx,natbib}
\usepackage{booktabs} 

\usepackage{subfigure}
\def\lapp{\ifmmode\stackrel{<}{_{\sim}}\else$\stackrel{<}{_{\sim}}$\fi}
\def\gapp{\ifmmode\stackrel{>}{_{\sim}}\else$\stackrel{>}{_{\sim}}$\fi}

\newcommand{\degrees}{^{\circ}}
\newcommand*\mean[1]{\overline{#1}}
\begin{document}
\pagenumbering{arabic}
\title{A Local Universe Host for the Repeating Fast Radio Burst FRB 20181030A }

\newcommand{\mcgillphysics}{Department of Physics, McGill University, 3600 rue University, Montr\'eal, QC H3A 2T8, Canada}
\newcommand{\msi}{McGill Space Institute, McGill University, 3550 rue University, Montr\'eal, QC H3A 2A7, Canada}
\newcommand{\wvuphysics}{Department of Physics and Astronomy, West Virginia University, P.O. Box 6315, Morgantown, WV 26506, USA}
\newcommand{\wvugws}{Center for Gravitational Waves and Cosmology, West Virginia University, Chestnut Ridge Research Building, Morgantown, WV 26505, USA}
\newcommand{\uoftphysics}{Department of Physics, University of Toronto, 60 St. George Street, Toronto, ON M5S 1A7, Canada}
\newcommand{\cita}{Canadian Institute for Theoretical Astrophysics, 60 St. George Street, Toronto, ON M5S 3H8, Canada}
\newcommand{\dunlapinstitute}{Dunlap Institute for Astronomy \& Astrophysics, University of Toronto, 50 St. George Street, Toronto, ON M5S 3H4, Canada}
\newcommand{\dunlapdep}{David A. Dunlap Department of Astronomy \& Astrophysics, University of Toronto, 50 St. George Street, Toronto, ON M5S 3H4, Canada}

\newcommand{\mitkavli}{MIT Kavli Institute for Astrophysics and Space Research, Massachusetts Institute of Technology, 77 Massachusetts Ave, Cambridge, MA 02139, USA}
\newcommand{\mitphysics}{Department of Physics, Massachusetts Institute of Technology, 77 Massachusetts Ave, Cambridge, MA 02139, USA}
\newcommand{\ubc}{Dept. of Physics and Astronomy, 6224 Agricultural Road, Vancouver, BC V6T 1Z1 Canada}
\newcommand{\sidrat}{Sidrat Research, PO Box 73527 RPO Wychwood, Toronto, Ontario, M6C 4A7, Canada}
\newcommand{\perimeter}{Perimeter Institute for Theoretical Physics, 31 Caroline Street N, Waterloo ON N2L 2Y5 Canada}
\newcommand{\tata}{Department of Astronomy and Astrophysics, Tata Institute of Fundamental Research, Mumbai, 400005, India}
\newcommand{\ncra}{National Centre for Radio Astrophysics, Post Bag 3, Ganeshkhind, Pune, 411007, India}
\newcommand{\drao}{Dominion Radio Astrophysical Observatory, Herzberg Astronomy \& Astrophysics Research Centre, National Research Council Canada, PO Box 248, Penticton, BC V2A 6J9, Canada}

\newcommand{\waterloo}{Department of Physics and Astronomy, University of Waterloo, Waterloo, ON N2L 3G1, Canada}
\newcommand{\maxico}{Instituto de Astronomía, Universidad Nacional Autónoma de México, Apdo. Postal 877, Ensenada, Baja California 22800, México}
\newcommand{\ioffe}{Ioffe Institute, 26 Politekhnicheskaya st., St. Petersburg 194021, Russia}
\newcommand{\maxicoinstitute}{Instituto Nacional de Astrof{\'\i}sica, \'Optica y Electr\'onica, Luis Enrique Erro 1, Tonantzintla 72840, Puebla, Mexico}

\author[0000-0002-3615-3514]{M.~Bhardwaj}
\affiliation{\mcgillphysics}
\affiliation{\msi}

\author[0000-0002-8139-8414]{A.~Yu.~Kirichenko}
\affiliation{\maxico}
\affiliation{\ioffe}

\author[0000-0002-2551-7554]{D.~Michilli}
\affiliation{\mcgillphysics}
\affiliation{\msi}

\author[0000-0002-2551-7554]{Y.~D.~Mayya}
\affiliation{\maxicoinstitute}

\author[0000-0001-9345-0307]{V.~M.~Kaspi}
\affiliation{\mcgillphysics}
\affiliation{\msi}

\author[0000-0002-3382-9558]{ B.~M.~Gaensler}
\affiliation{\dunlapinstitute}
\affiliation{\dunlapdep}

\author[0000-0003-1842-6096]{M.~Rahman}
\affiliation{\sidrat}

\author[0000-0003-2548-2926]{S.~P.~Tendulkar}
\affiliation{\tata}
\affiliation{\ncra}


\author[0000-0001-8384-5049]{E. Fonseca}
\affiliation{\mcgillphysics}
\affiliation{\msi}

\author[0000-0003-3059-6223]{Alexander Josephy}
\affiliation{\mcgillphysics}
\affiliation{\msi}

\author[0000-0002-4209-7408]{C. Leung}
\affiliation{\mitkavli}
\affiliation{\mitphysics}

\author[0000-0003-2095-0380]{Marcus Merryfield}
\affiliation{\mcgillphysics}
\affiliation{\msi}

\author[0000-0002-9822-8008]{Emily Petroff}
\affiliation{\mcgillphysics}
\affiliation{\msi}
\affiliation{Anton Pannekoek Institute for Astronomy, University of Amsterdam, Science Park 904, 1098 XH Amsterdam, The Netherlands}
\affiliation{Veni Fellow}

\author[0000-0002-4795-697X]{Z.~Pleunis}
\affiliation{\mcgillphysics}
\affiliation{\msi}

\author[0000-0001-5504-229X]{Pranav Sanghavi}
  \affiliation{Lane Department of Computer Science and Electrical Engineering, 1220 Evansdale Drive, PO Box 6109, Morgantown, WV 26506, USA}
  \affiliation{Center for Gravitational Waves and Cosmology, West Virginia University, Chestnut Ridge Research Building, Morgantown, WV 26505, USA}

\author[0000-0002-7374-7119]{P.~Scholz}
\affiliation{\dunlapinstitute}

\author[0000-0002-6823-2073]{K.~Shin}
\affiliation{\mitkavli}
\affiliation{\mitphysics}

\author[0000-0002-2088-3125]{Kendrick M.~Smith}
  \affiliation{Perimeter Institute for Theoretical Physics, 31 Caroline Street N, Waterloo, ON N25 2YL, Canada}

\author[0000-0001-9784-8670]{I.~H.~Stairs}
\affiliation{\ubc}

\correspondingauthor{Mohit Bhardwaj}
\email{mohit.bhardwaj@mail.mcgill.ca}
\begin{abstract}
We report on the host association of FRB 20181030A, a repeating fast radio burst (FRB) with a low dispersion measure (DM, 103.5 pc cm$^{-3}$) discovered by \cite{abb+19c}. Using baseband voltage data saved for its repeat bursts, we localize the FRB to a sky area of 5.3 sq. arcmin (90\% confidence). Within the FRB localization region, we identify NGC 3252 as the most promising host, with an estimated chance coincidence probability $< 2.5 \times 10^{-3}$. Moreover, we do not find any other galaxy with M$_{r} < -15$ AB mag within the localization region to the maximum estimated FRB redshift of 0.05. This rules out a dwarf host 5 times less luminous than any FRB host discovered to date. NGC 3252 is a star-forming spiral galaxy, and at a distance of $\approx$ 20 Mpc, it is one of the closest FRB hosts discovered thus far. From our archival radio data search, we estimate a 3$\sigma$ upper limit on the luminosity of a persistent compact radio source (source size $<$ 0.3 kpc at 20 Mpc) at 3 GHz to be ${\rm 2 \times 10^{26} erg~s^{-1} Hz^{-1}}$, at least 1500 times smaller than that of the FRB 20121102A persistent radio source. We also argue that a population of young millisecond magnetars alone cannot explain the observed volumetric rate of repeating FRBs. Finally, FRB 20181030A is a promising source for constraining FRB emission models due to its proximity, and we strongly encourage its multi-wavelength follow-up.
\end{abstract}


\section{introduction}

Fast radio bursts (FRBs) are enigmatic radio pulses of high brightness temperature ($\sim$ 10$^{35}$ K) and millisecond duration \citep{lbm+07,tsb+13}. In spite of the fact that more than 500 FRBs have been discovered to date\footnote{For a complete list of known FRBs, see \url{https://www.herta-experiment.org/frbstats/} or the TNS \citep{2020TNSAN..70....1Y}.}, their nature continues to be a subject of intense debate, owing 
in part to a limited sample of localized FRBs. Furthermore, the FRBs exhibit a diverse range of phenomenology: most of the discovered sources are apparently non-repeating, but a small fraction are observed to repeat. Among the repeating FRBs, two thus far have shown evidence of
periodic repetitions \citep{Amiri:2020gno,Rajwade:2020uat,2021MNRAS.500..448C}. As a result,  a plethora of theories has been proposed to explain the FRB sources' disparate behavior \citep[see][for a catalog of proposed models]{pww+18}.  

To unveil the nature of FRB sources, detailed studies of FRB hosts and their local environments are a promising way forward \citep{nicholl2017,li2020}. 
Currently, only 15 published FRBs have been sufficiently well localized on the sky to allow their host galaxies to be identified.\footnote{See \url{http://frbhosts.org/} (visited on 01/07/2021); \cite{heintz2020host}.}
All localized FRBs except FRB 20200120E in M81 \citep{bhardwaj2021} 
are located at redshifts ranging from 0.03 to 0.66 where the detailed study of the FRB local environment is 
limited by the sensitivity of current telescopes. Additionally, FRBs so far are only observed at radio wavelengths with no convincing afterglow emission reported to date. However, X-ray emission contemporaneous with FRB-like radio bursts were detected from SGR 1935+2154, which suggests that at least some FRBs could have prompt X-ray counterparts \citep{2020SGR,Bochenek2020,mereghetti2020integral,li2020identification,ridnaia2020peculiar}. Such X-ray emission currently can only be detected for nearby FRBs
\citep[$<$ 50 Mpc;][]{scholz2020simultaneous}. 

Recently, \cite{bhardwaj2021} reported the discovery of an FRB with the lowest DM observed to date, FRB 20200120E, 
located toward the outskirts of the nearby spiral galaxy M81. Using the European Very-long-baseline interferometry Network (EVN), \cite{Kristan2021arXiv} localized the FRB with sub-arcsecond precision to an M81 globular cluster. At a distance of 3.6 Mpc, FRB 20200120E is an excellent candidate for multi-wavelength observations that could strongly constrain the nature of the FRB progenitor.

Here we report the identification of the most likely host for FRB 20181030A\footnote{Formerly named as FRB181030.J1054+73.}, a repeating FRB first reported by \cite{abb+19c}. 
Though its DM is only 103.5 pc cm$^{-3}$, this is significantly larger than the expected contribution in this direction from the Milky Way disk ($\sim$ 33$-$41 pc cm$^{-3}$).
\cite{abb+19c} did not find any Galactic ionized and/or star-forming region in the direction of FRB 20181030A. As a result, they concluded that the FRB should have a nearby extragalactic host. However, due to insufficiently precise localization of the FRB reported by \cite{abb+19c}, they could not make any firm association with a host.  Since that report, CHIME/FRB has detected seven more bursts from the FRB (see Table \ref{ta:baseband_bursts}).\footnote{For a complete list, check \url{http://chime-frb.ca/repeaters/FRB20181030A} (visited on 1/07/2021).\label{chimewebsite} } For several of the FRB repeat bursts, raw voltage data were acquired, enabling localization of the FRB to a few arcminute precision, an improvement of over a factor of 200 in localization area.
Within this localization region, we identify a local Universe spiral galaxy, NGC 3252 \citep{huchra1983ApJS}, as its most likely host.

The paper is organized as follows: In \S\ref{sec:obs}, we describe our search for the host of FRB20181030A. From the low chance coincidence probability (\S\ref{sec:host-search}) and absence of any other viable host candidates in the FRB localization region (\S\ref{sec:maxz}), we argue that NGC 3252 is a promising host for the FRB. We estimate notable physical properties of NGC 3252 in \S\ref{sec:physical-properties} and then discuss our archival multi-wavelength data search to identify any FRB plausible counterpart in \S\ref{sec:multi-wavelength}. In \S\ref{sec:discussion}, we discuss 
implications of this discovery, and 
conclude in \S\ref{sec:conclusion}.

\section{Observations}
\label{sec:obs}

 The Canadian Hydrogen Intensity Mapping Experiment Fast Radio Burst Project (CHIME/FRB) \citep{abb+2018ApJ} first discovered two bursts from FRB 20181030A on 2018 October 30
\citep{abb+19c}. The source's DM is larger than the predicted Galactic contribution in the FRB sight-line (See Table \ref{tab:params}). 
After subtracting DM contributions from the Milky Way disk and halo, as shown in Table \ref{tab:params}, the DM-excess of the FRB is $\sim30-40~$pc cm$^{-3}$. Using the average Macquart relation \citep[Equation 2 in][]{macquart2020census}, we estimate the redshift of the FRB to be $\sim 0.03-0.04$ assuming negligible host DM contribution. This suggests close proximity of the FRB host ($\lesssim$ 200 Mpc). 
As of 2021 July 1, seven more bursts have been detected from the FRB{\textsuperscript{\ref{chimewebsite}}}. 
Interestingly, all seven bursts were clustered in two different epochs on 2020 January 22, separated by $\approx$ 12 hours. This suggests a highly non-Poissonian waiting time distribution for the FRB bursts. Fortunately, four FRB 20181030A bursts have baseband data saved by the CHIME/FRB baseband system (bursts with reported DM${\rm_{bb}}$ values in Table \ref{ta:baseband_bursts}).

\begin{figure}[ht]
\includegraphics[width=.90\linewidth]{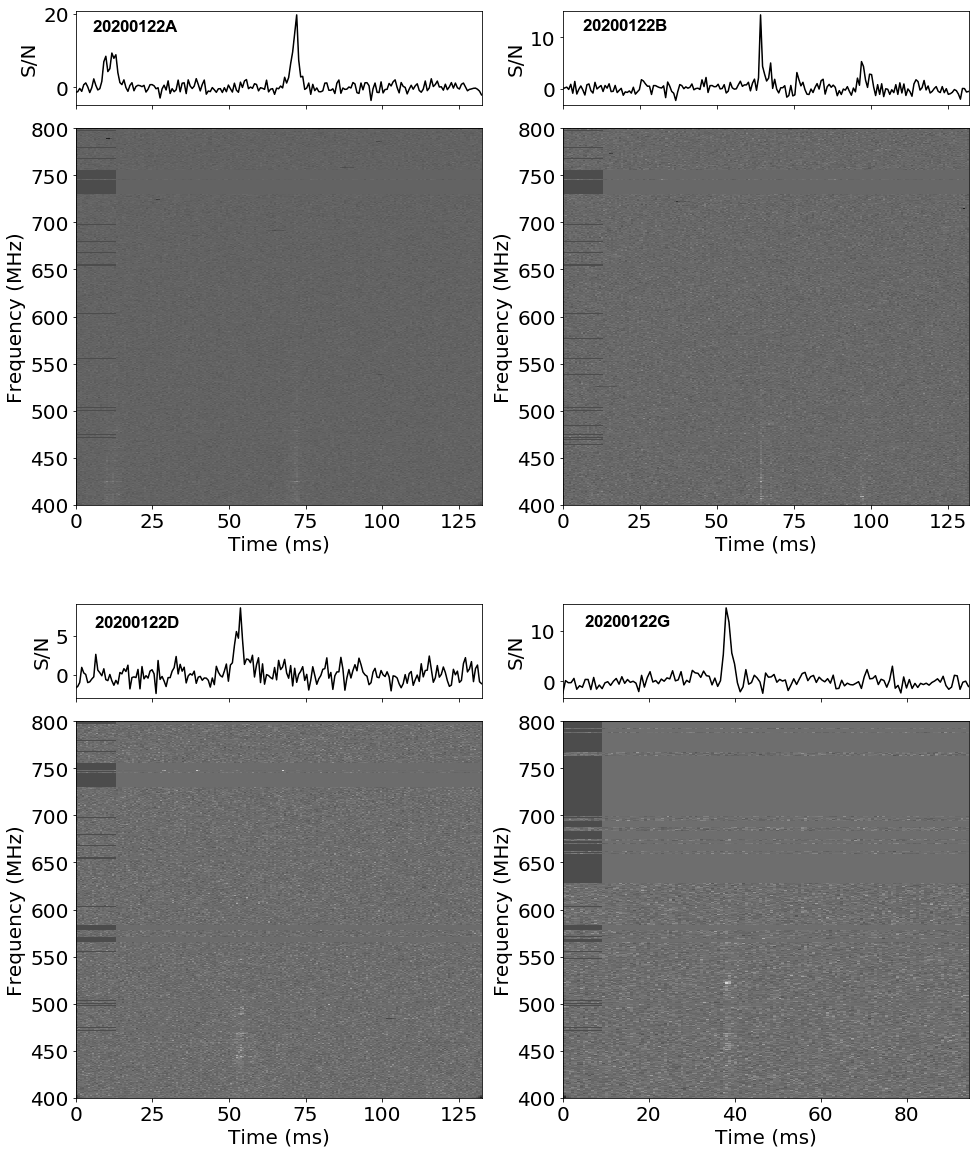}
\caption{Frequency versus time (``waterfall") plots of the four dedispersed bursts detected from FRB 20181030A with saved baseband data. See Table \ref{ta:baseband_bursts} for their major burst properties. The waterfall plots are binned to have temporal resolution 0.655 ms and spectral resolution 0.391 MHz. Dark grey lines represent bad frequency channels that were flagged in this analysis. Note that FRBs 20200122A and 20200122B show sub-bursts separated by $\sim$ 60 ms and 30 ms, respectively. Detailed analysis of these sub-bursts will be reported elsewhere.
}
\label{fig:waterfall}
\end{figure}

The baseband system of CHIME/FRB stores $\sim 100$\,ms of channelized voltages around signals of interest \citep{abb+2018ApJ}. 
We have developed a pipeline to automatically process such baseband data and localize a burst on the sky with a precision of $\sim\frac{8}{\text{S/N}}$\,arcmin \citep{Daniele2020}.
This is achieved by mapping the signal strength with a grid of largely overlapping beams around an initial guess. 
The resulting S/N measured in each beam is fitted with a mathematical model of the formed beam of the telescope.
Systematic effects have been corrected by using a sample of sources with known positions. In this case, we used the baseband data of four detected bursts of FRB 20181030A, FRBs 20200122A, 20200122B, 20200122D, and 20200122G, to estimate the localization region of the FRB. The dedispersed baseband data waterfall plots and major 
characteristics
of the four FRB bursts are shown in Figure \ref{fig:waterfall} and Table \ref{ta:baseband_bursts}, respectively. Other burst properties, such as fluence and flux density, along with a detailed description of both the intensity and baseband data analysis of all newly discovered FRB 20181030A bursts will be presented elsewhere. Moreover, the available data is insufficient to estimate meaningful constraints on the FRB's periodicity. As the reported baseband localization uncertainties are statistical in nature \citep{Daniele2020}, we combined the localization regions of the four FRB bursts using a weighted average with inverse variance weights and
 localized the FRB to a sky area of $\approx$ 5.3 arcmin$^{2}$ (90\% confidence region; see Table \ref{tab:params}). Next, we use the baseband localization region of FRB 20181030A to search for a potential host galaxy.  

\begin{table}[ht]
\begin{center}
\caption{Properties of the bursts from FRB 20181030A.}
\hspace{-1.in}
\begin{tabular}{cccccc} 
\hline
TNS Name & MJD & Arrival Time$^a$& S/N$^{b}$ & DM$_\mathrm{bb}^c$ & DM$^d$\\
 &  & (UTC @ 400 MHz) &  & (pc~cm$^{-3}$) & (pc~cm$^{-3}$) \\
\hline
FRB 20181030A$^e$ & 58421 & 04:13:13.1758(8) &  32.5 & $-$ & 103.5 $\pm$ 0.7\\
FRB 20181030B$^e$ & 58421 & 04:16:21.6419(14) & 17.1 & $-$ & 103.5 $\pm$ 0.3\\

FRB 20200122A & 58870 & 10:20:32.5805(3) & 13.9 & 103.53 $\pm$ 0.02 & 103.40 $\pm$ 0.14\\
FRB 20200122B & 58870 & 10:27:00.4412(3) & 17.3 & 103.49 $\pm$ 0.02 & 103.47 $\pm$ 0.08\\

FRB 20200122C & 58870 & 10:28:20(1) & 8.3 & $-$ &  103.1 $\pm$ 1.2 \\

FRB 20200122D & 58870 & 22:09:30.8575(3) & 13.1 & 103.58 $\pm$ 0.19 & 103.7 $\pm$ 0.4\\

FRB 20200122E & 58870 & 22:09:52(1)  & 10.4 & $-$ & 103.27 $\pm$ 0.13\\
 FRB 20200122F & 58870 & 22:22:21(1)  & 8.9 & $-$ & 103.7 $\pm$ 0.7 \\

FRB 20200122G & 58870 & 22:23:20.3080(3) & 10.5 & 103.57 $\pm$ 0.10 & 103.7 $\pm$ 0.5\\
\hline
\end{tabular}
\label{ta:baseband_bursts}
\end{center}
$^a$ All burst times of arrival are topocentric. For FRBs 20200122C, 20200122E, and 20200122F, the arrival times are reported by the CHIME/FRB real-time pipeline \citep{abb+2018ApJ}. For FRBs 20200122A, 20200122B, 20200122D, and 20200122G, the  arrival times are estimated by the baseband pipeline \citep{Daniele2020}. Finally, the arrival times of FRBs 20181030A and 20181030B are taken from \cite{abb+19c}.\\
$^b$ For all except FRBs 20181030A and 20181030B, band-averaged signal-to-noise (S/N) ratios are estimated by the CHIME/FRB real-time pipeline.\\
$^c$ S/N-optimized DM for the bursts detected in the baseband data.\\
$^d$ S/N-optimized DM from intensity data \citep[see][]{abb+19c,fab+20}.\\
$^e$ Data from \cite{abb+19c}.\\
\end{table}





\begin{table}[ht]
\begin{center}
\caption{Major Observables of FRB 20181030A.}
\begin{tabular}{@{} *2l @{}}\toprule
\textbf{Parameter} & \textbf{Value}\\\midrule
R.A.(J2000)$^a$ & $\rm{10^{h}34^{m}20^{s}.1 \pm 30^{s}.6}$ \\ 
Dec. (J2000)$^a$ & $\rm{73\degrees45\arcmin05\arcsec \pm 47\arcsec}$ \\
$l,b$ & 134.$\degrees$81, +40.$\degrees$06 \\
DM$^b$ &  103.5~$\pm$~0.3  pc cm$^{-3}$\\
DM$_{\mathrm{MW, NE2001}}^c$ & 41 pc cm$^{-3}$ \\
DM$_{\mathrm{MW, YMW16}}^c$ & 33 pc cm$^{-3}$ \\
DM$_{\mathrm{MW, halo}}^d$ & 30 pc cm$^{-3}$ \\
Max. distance$^{e}$ & $\lesssim$ 225 Mpc \\
 \hline
\end{tabular}
\label{tab:params}
\end{center}
$^a$ The 90\% confidence localization region of the FRB.

$^b$ From \cite{andersen2019chime}.

$^c$ Maximum DM model prediction along this line-of-sight for the NE2001 \citep{cordes2002ne2001} and YMW16 \citep{yao2017new} Galactic electron density distribution models. 

$^d$Fiducial Milky Way halo prediction from the \cite{dolag2015constraints} hydrodynamic simulation and \cite{yamasaki2020galactic} Milky Way Halo model.

$^e$ Corresponds to the maximum redshift of 0.05 (see \S\ref{sec:maxz}).
 
\end{table}

\begin{figure}[ht]
\includegraphics[width=.95\linewidth]{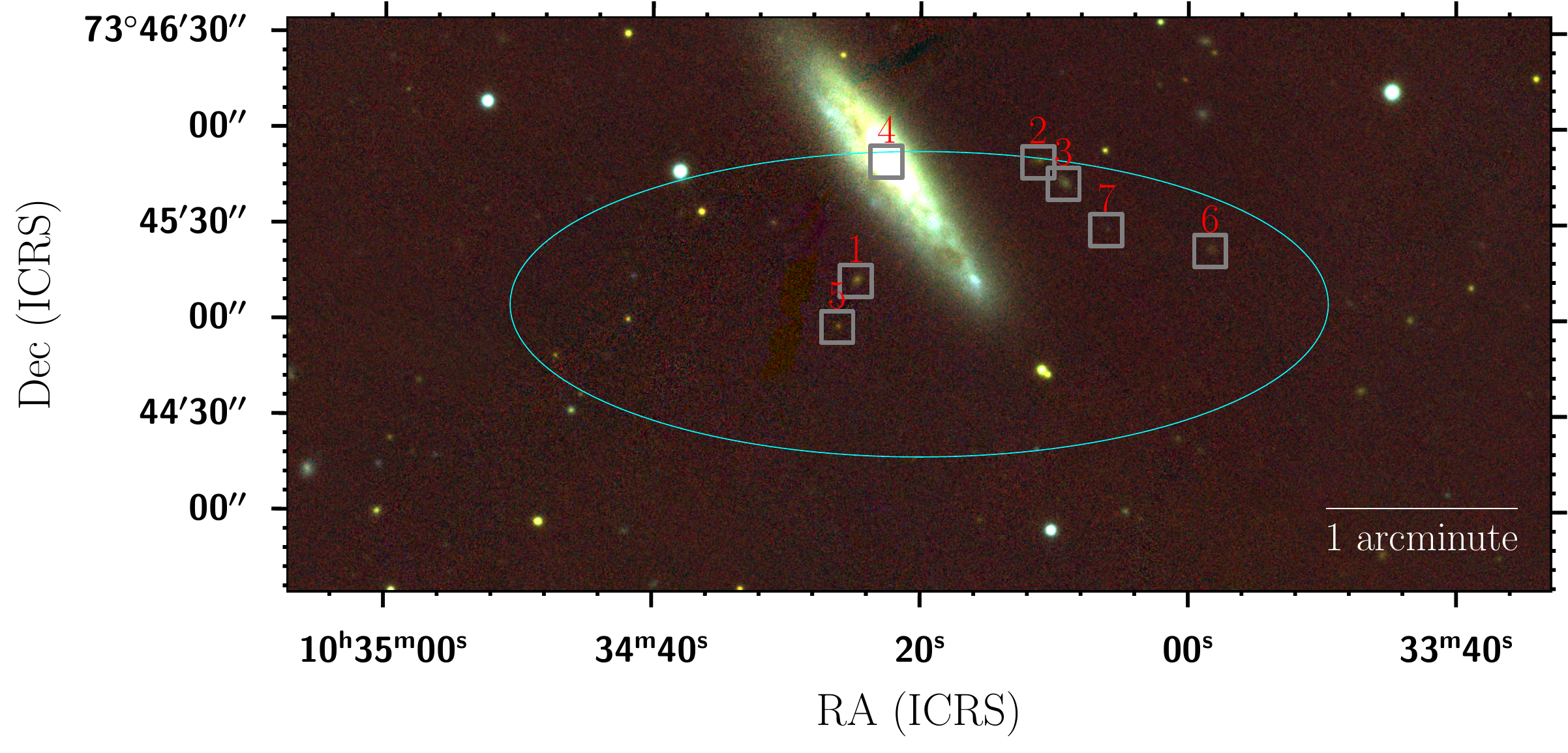}
\caption{Pan-STARRS RGB-image of the FRB 20181030A 90\% localization region (cyan ellipse). Grey boxes show the locations of 7 host galaxy candidates within the localization region (See Table \ref{Tab2}); Source 4 is NGC 3252 at z = 0.0039, the most promising host galaxy of the FRB. 
}
\label{fig:frbfov}
\end{figure}


\subsection{Host galaxy search}
\label{sec:host-search}

 First, we argue below that 
 the FRB is unlikely to be Galactic in origin. As noted by \cite{abb+19c}, there is no catalogued Galactic ionized region, satellite galaxy, or globular cluster in the direction of the FRB that could contribute to the FRB DM. Moreover, \cite{ocker2020ApJ} estimated the mean DM through the Milky Way's warm ionized medium at large distances from the Galactic plane  ($z >$ 2 kpc), $\overline{{\rm{DM} \sin|b|}}$ = 23.0 $\pm$ 2.5 pc cm$^{-3}$. At the FRB's Galactic latitude ($b = 40 \degrees$), it would give a mean Galactic DM of $\approx 36 \pm 5$~pc~cm$^{-3}$. This agrees well with the prediction of the two Galactic DM models. The Milky Way halo DM contribution, DM$_{\mathrm{halo}}$, on the other hand, is poorly constrained. Recently, \cite{Kristan2021arXiv} estimated the Milky Way halo contribution in the direction of FRB 20200120E to be $\lesssim$ 40 pc cm$^{-3}$. If this is also true for the FRB 20181030A sight-line, the FRB would be clearly extragalactic in origin. However, the halo may be clumpy \citep{Kaaret2020}, so it may still be possible to have significant variations in DM$_{\mathrm{halo}}$ along different sight-lines. Using the same argument as asserted by \cite{bhardwaj2021}, an FRB with a DM-excess of $\sim 70$ pc cm$^{-3}$, if Galactic, would require a very distant ($\gtrsim$ 100 kpc) and unusually energetic neutron star as its source. As discussed below, we have found an extragalactic host with a low chance coincidence probability. Therefore,
Occam's razor argues for the extragalactic association.

Next, we searched the NASA Extragalactic Database (NED) for catalogued galaxies within the FRB 90\% confidence localization region and found only one galaxy, NGC 3252, with a redshift (z) of 0.00385(2) \citep{2014MNRAS.443.1044M}. NGC 3252 is a bright (m$_{\text{r}}$ = 12.58 AB mag) Scd Hubble-type edge-on spiral galaxy \citep{de1991third} at a luminosity distance of $\approx 20$ Mpc.
In Figure \ref{fig:frbfov}, we plotted the FRB localization region over a Pan-STARRS RGB image made using Pan-STARRS's g-band (B:blue), r-band (G:green), and z-band (R:red) data. In the Figure, NGC 3252 is the most prominent galaxy. Note that NED does not provide the depth of completeness of catalogued galaxies in their search results. Therefore, in \S\ref{sec:maxz}, we describe our search of dwarf galaxies within the FRB localization region.

\begin{figure}[ht]
\includegraphics[width=.95\linewidth]{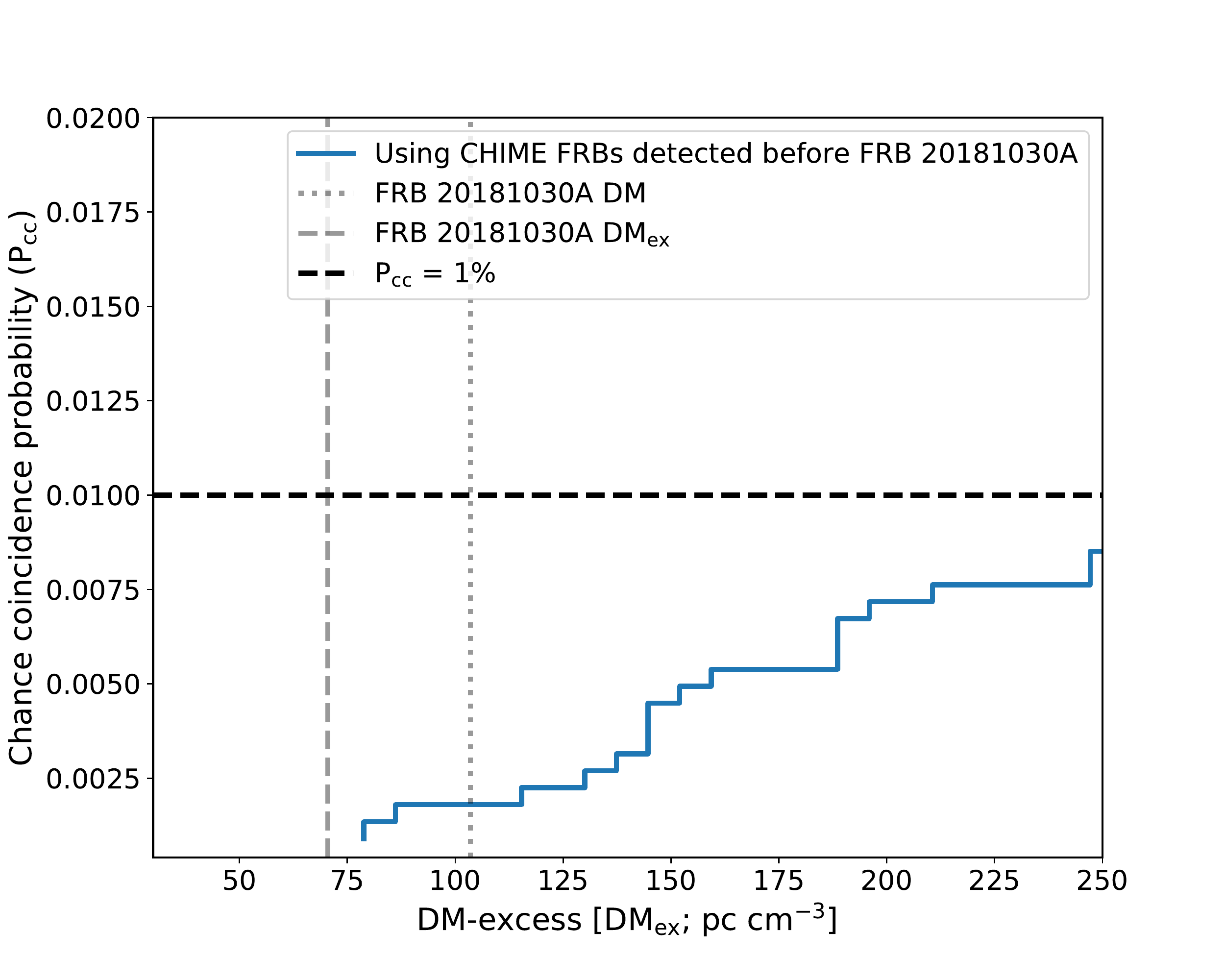}
\caption{The chance coincidence probability of finding an NGC 3252-like galaxy as a function of the DM-excess (DM$\rm{_{ex}}$) of CHIME FRBs detected before the first burst of FRB20181030A (see \S \ref{sec:host-search}). As discussed by \cite{bhardwaj2021}, the latter step takes into account the look-elsewhere effect.
Given the DM of FRB 20181030A, 103.5 pc cm$^{-3}$, P$\mathrm{_{cc}} <$  0.0025.
}
\label{fig:fig}
\end{figure}

We now estimate the chance coincidence probability (P$_{\mathrm{cc}}$) of finding an NGC 3252-like bright galaxy close to the FRB localization region. 
Briefly, we assume a Poisson distribution of galaxies across the sky and calculate the probability of finding one or more galaxies with m$_{\text{r}}$ smaller than or equal to that of NGC 3252 (12.79 AB mag; without correcting for the Galactic extinction) by chance close to the FRB 90\%-confidence localization region (5.3 acmin$^{2}$). Using the areal number density of NGC 3252-like or brighter galaxies, $n\rm{(m_{r} \leq 12.79)}$ = 0.2 deg$^{-2}$ from \cite{driver2016measurements}, we estimate P$_{\mathrm{cc}} = 4.5 \times 10^{-4}$. However, as the presence of NGC 3252 is inferred post-hoc, we have corrected the P$\rm{_{cc}}$ to account for the problem of multiple testing (also known as the look-elsewhere effect) using the method 
described by \cite{bhardwaj2021}. After considering all CHIME FRBs that were discovered before the first detected burst of FRB 20181030A and have the DM-excess $\leq$ 103.5 pc cm$^{-3}$ (see Figure \ref{fig:fig}), we estimate the P$_{\rm{cc}}$  to be $<$ 0.0025. 


We should point out, however, that our chance coincidence analysis favors brighter galaxies over fainter ones because the latter are more abundant and therefore more likely to be found in the FRB localization region by chance. Therefore, in the next section, we searched for faint galaxies within the FRB localization region. 

\subsection{A dwarf host of FRB 20181030A?}
\label{sec:maxz}
In order to check if there exists any plausible dwarf galaxy within the FRB localization, we first estimated the maximum redshift of the FRB 20181030A host by performing a Markov Chain Monte Carlo (MCMC) simulation, which is discussed in Appendix \ref{app:mcmc}. Using the priors shown in Table \ref{tab:mcmc-priors} and an \texttt{emcee} based MCMC inference framework \citep{fm2013}, we computed a one-sided 95\% Bayesian credible upper limit on the FRB host redshift (z$_{\mathrm{max}}$) = 0.05 from the marginalized host redshift posterior. 

There are a few factors that make our maximum redshift estimate conservative. If the FRB host lies beyond NGC 3252, the FRB sight-line would traverse the NGC 3252 halo with a projected offset $\leq$ 14 kpc. Using the stellar mass of NGC 3252 from Table \ref{tab:host-properties}, we estimated its halo mass to be 1.9 $\times 10^{11}$ M$_{\odot}$ using stellar mass to halo mass relation using Equation 2 from \cite{moster2013} and the NFW profile halo concentration factor = 9.4 using Equation 24 from  \cite{klypin2016}. At a projected offset of 14 kpc, using the method described by \cite{bhardwaj2021}, we estimate the DM contribution of the NGC 3252 halo $\approx$ 15 or 30 pc cm$^{-3}$ for baryon fractions 0.4 and 0.75, respectively, using the halo density profile from \cite{maller2004multiphase} \citep[the profile that predicted the lowest M81 halo DM in Figure 4 of ][]{bhardwaj2021}. The former baryon fraction value is the minimum that \cite{hafen2019origins} found in the Feedback In Realistic Environments (FIRE) simulation for a halo of mass  $\sim$ 10$^{11}$ M$_{\odot}$). The latter value, i.e., 0.75, is estimated assuming $\approx$ 25\% of the baryons exist in the galaxy as the interstellar medium (ISM), stars, and compact remnants \citep{fukugita1998cosmic}. Moreover, NGC 3252 is a part of a galaxy group with the dynamic group mass = 1.2 $\times 10^{12}$ M$_{\odot}$ \citep{Kourkchi2017}. In addition to this, the FRB sight-line intersects several other galaxy groups that are located within z = z$_{\mathrm{max}}$ \citep{tempel2016,lim2017}. All these would contribute to the observed FRB DM and consequently, if accounted for, would reduce our maximum redshift estimate considerably. 

Note that an FRB 20121102A-like star-forming dwarf galaxy \citep[$\mathrm{M_{r}} = -17$ 
AB mag;][]{tendulkar2017host}, the faintest FRB host discovered to date, if located at z$_{\mathrm{max}}$, would have r-band magnitude $\approx~$19.8 AB mag. 
Fortunately, the FRB field-of-view is imaged by the Dark Energy Spectroscopic Instrument (DESI) Legacy Imaging Survey \citep{DESI-2019}  with an r-band depth $\approx$ 24 AB mag (5$\sigma$). Using the DESI data, \cite{Zou2019ApJS} estimated photometric redshifts of all the identified galaxies with a 5$\sigma$ r-band completeness limit of 23.6 AB mag. 
However, this limit is likely not complete for low surface brightness (LSB) galaxies. For the DESI Legacy Imaging Survey, the average r-band surface brightness limit is $\sim$ 26 mag arcsec$^{-2}$ \citep{,DESI-2019,tanoglidis2021ApJS,arora2021}.
At z = z$_{\mathrm{max}}$, an M$_{r} = -17$ AB mag LSB galaxy of effective radius $\sim$ 1-3 kpc \citep{Greco2018ApJ}
and uniform surface brightness $=$ 26 mag arcsec$^{-2}$
should be detected in the DESI data as an m$_{r}$ $\lesssim$ 22 AB mag source. More importantly, the m$_{r}$ $\lesssim$ 22 AB mag limit is sensitive to detect a dwarf host 5 times less luminous than any FRB host discovered to date (M$_{r} = -15$ AB mag). Given this constraint, we selected seven galaxies from \cite{Zhou2020ApJ}, including NGC 3252, which have m$_{r}$ $\leq$ 22 AB mag and are located within the FRB 90\% confidence localization region (shown as a cyan ellipse in Figure \ref{fig:frbfov})  and estimated their spectroscopic redshifts using the 10.4-m Gran Telescopio Canarias (GTC) (see \S\ref{sec:GTC}).  

\subsection{GTC observations and analysis}
\label{sec:GTC}

In this section, we describe our GTC observations of the seven plausible host candidates. As will be shown later, only NGC 3252 satisfies the z$_{\rm max}$ constraint, and hence, the most plausible host of the FRB.
\subsubsection{Observations}

Observations of the galaxies identified within the FRB 20181030A 90\% localization region were performed with the the Optical System for Imaging and low-intermediate Resolution
Integrated Spectroscopy (OSIRIS\footnote{\url{http://www.gtc.iac.es/instruments/osiris/}})  
at the GTC. The OSIRIS detector consists of two CCDs and provides a field of view (FoV) of 7.8$\arcmin \times$~7.8$\arcmin$   
with a pixel scale of 0.254$\arcsec$. The data were obtained during four observing runs in October 2020 and May 2021. The observing blocks corresponding to the first two runs were executed under Director's Discretionary time. A summary of the observations is given in Table \ref{log}.  

We obtained long-slit spectra of the 
NGC~3252 using the R1000B grism that covers the spectral range from 3700 to 7500 \AA, with the 1.2$\arcsec$ slit width, providing a spectral resolution of about 9 Å. The slit was placed to pass through the major 
axis of the galaxy at a PA=37.31$^\circ$, which is shown in Figure~\ref{fig:longslit-ngc3252}. 

To perform simultaneous observations of the other six host galaxy candidates in the localization region (see Table \ref{Tab2}), we utilized the OSIRIS MOS (multi-object spectroscopy) mode. The mask for the MOS observations was designed with the OSIRIS Mask Designer Tool \citep{mask1,mask2} 
using the catalog coordinates of the galaxies and a set of five fiducial stars. 
The observations were performed with the R500B and R500R grisms that
cover the spectral ranges 3600$-$7200 Å and 4800$-$10000 Å, respectively. 
For the target galaxies we designed rectangular slitlets with length varying between 4.5$\arcsec$ and 10$\arcsec$ and a width of 1.5$\arcsec$ each. Two additional slitlets covered source-free regions for sky subtraction. The spectral resolution of the R500B and R500R data is $\sim$21 Å and $\sim$27 Å, respectively. 

\begin{table}[ht]
\caption{Log of the GTC/OSIRIS long-slit and MOS spectroscopic observations of the FRB 20181030A 90\% localization region.}
\begin{center}
\hspace{-1.in}
\begin{tabular}{@{} *9l @{}}
\toprule
\textbf{Program} & \textbf{Date} & \textbf{Mode} & \textbf{Grism}& \textbf{Position} & \textbf{Exposure} & \textbf{Seeing} & \textbf{Airmass} & \textbf{Night}   \\
 & & & & \textbf{Angle} & \textbf{Time}  &  & & \\ 
 \midrule 
  GTC04-20BDDT & 24/10/2020 & long-slit & R1000B & 37.31$^\circ$ & 4 $\times$ 60 s & 1.2$\arcsec$ & 1.59 & Dark     \\
  GTC04-20BDDT &  26/10/2020 & MOS & R500R & 0 & 8 $\times$ 700 s & 1.5$\arcsec$ & 1.75 & Dark  \\
  GTC18-21AMEX & 04/05/2021 & MOS & R500B & 0 & 3 $\times$ 1200 s & 0.9$\arcsec$  & 1.68  & Dark   \\
  GTC18-21AMEX & 15/05/2021 & MOS & R500B & 0 & 3 $\times$ 1200 s & 1.0$\arcsec$  & 1.56  & Dark   \\\bottomrule 
 \hline
\end{tabular}
\label{log}
\end{center}
\end{table}

\subsubsection{Data reduction}
\label{sec:mos}
The OSIRIS MOS and long-slit spectra were reduced using the GTCMOS pipeline \citep{gtcmos} and standard IRAF routines \citep{tody1986iraf,tody1993iraf}. All spectra were bias-subtracted, 
and flat-fielded using the set of corresponding images taken during the same observing nights.
%
For flux calibration we used spectrophotometric standards Feige 110, GD153 and Ross 640 \citep{standard, std3, std2} observed during the same nights as the targets. 
A set of arc-lamp spectra of Ne, Hg and Ar was used for wavelength calibration. 
The rms errors of the resulting solutions were $<$0.5\AA\ for the R1000B grating and $<$2 Å for the R500R and R500B gratings. 

\begin{table}[ht]
\caption{Galaxies identified within the FRB localization region with M$_{\rm r} < -15$ AB mag at z$_{\rm max}$ = 0.05.}
\begin{center}
\hspace{-1.in}
\resizebox{0.95\textwidth}{!}{ 
\begin{tabular}{@{} *6c @{}}
\toprule
\textbf{Number} & \textbf{R.A.} &\textbf{Dec.}& \textbf{DESI(r-band)}$^{a}$& \textbf{Identified lines} & \textbf{z$_{\mathrm{spec}}$}\\
&J2000& J2000& AB mag.& \\ \midrule 
1 & \rm{10$^{h}$34$^{m}$24.$^{s}$81} & 73$^{\degrees}$45$^{\arcmin}$12.$^{\arcsec}$81 & 19.69 & [OII], Ca doublet, G-band  & 0.460(1)   \\ 
2 & \rm{10$^{h}$34$^{m}$11.$^{s}$23} & 73$^{\degrees}$45$^{\arcmin}$49.$^{\arcsec}$23 & 19.89 & Ca doublet, G-band, Mg, Na & 0.455(2)\\ 
3 & \rm{10$^{h}$34$^{m}$9.$^{s}$33} & 73$^{\degrees}$45$^{\arcmin}$42.$^{\arcsec}$33 & 19.41 & [OII], Ca doublet, [OIII] doublet  &  0.276(2)   \\ 
\textbf{4}$^{b}$ & \textbf{10$^{h}$34$^{m}$22.$^{s}$56} & \textbf{73$^{\degrees}$45$^{\arcmin}$49.$^{\arcsec}$56} &  \textbf{12.58} & see text & \textbf{0.00385(2)}\\
5 & \rm{10$^{h}$34$^{m}$26.$^{s}$20} & 73$^{\degrees}$44$^{\arcmin}$57.$^{\arcsec}$20 & 21.61 & Ca doublet, G-band, Mg, Na & 0.645(1)     \\ 
6 & \rm{10$^{h}$33$^{m}$58.$^{s}$36} & 73$^{\degrees}$45$^{\arcmin}$21.$^{\arcsec}$36 & 20.76 & Ca doublet, G-band & 0.647(1)     \\    
7 & \rm{10$^{h}$34$^{m}$6.$^{s}$12} & 73$^{\degrees}$45$^{\arcmin}$28.$^{\arcsec}$12 & 21.67 &  [OII], Ca doublet & 0.563(2)    \\ 
 \hline
\end{tabular}
}
\end{center}

$^a$ The r-band magnitudes are corrected for the Milky Way extinction.\\
$^b$ Source 4 is NGC 3252, and at a spectroscopic redshift = 0.0039 (20 Mpc), it is the only galaxy in our list with the redshift $< \rm z_{max}$.\\
\label{Tab2}
\end{table}

\subsubsection{Multi-object Spectroscopy}

The resulting product of the pipeline contained 2-D calibrated spectra collected in all of the slitlets. We extracted each spectrum and subtracted the sky using the IRAF task $apall$. 
We utilized background from the source-free regions to subtract sky from the spectra obtained in the shortest slitlets. 
The lines identified for each galaxy and the corresponding average redshifts are presented in Table \ref{Tab2}. 
To confirm our redshift estimations, we used the Manual and Automatic Redshifting Software  \citep[MARZ,][]{Hinton} and compared the extracted spectra with the galaxy templates. In all cases the identified 
spectral lines (see Table \ref{Tab2}) have shown an agreement with the spectral features corresponding to early type absorption and intermediate type galaxy templates, confirming our estimations. 

Among the identified host galaxy candidates, only NGC 3252 has a spectroscopic redshift $< z_{\rm{max}}$. This makes NGC 3252 the only viable FRB host candidate among all the identified galaxies with m$_{\rm{r}} \leq 22$ AB mag. 
Note that blue star-forming dwarf galaxies have been proposed to host FRB progenitors \citep{metzger2017ApJ} via ``prompt"-formation-channels, such as superluminous supernovae and long gamma-ray bursts \citep{fruchter2006}. However, because of their highly dynamic and rich ISM, these galaxies are expected to contribute significantly to the FRB DM
\citep{Li2019ApJ}.
For instance, \cite{tendulkar2017host} estimated that the DM contribution of the FRB 20121102A host, a dwarf irregular star-forming galaxy, is $\sim 60 - 220$ pc cm$^{-3}$. 
Hence, together with the inference from \S\ref{sec:maxz}, the prospect of a host galaxy beyond NGC 3252 seems unlikely.
Lastly, in Table 5, we have listed the three galaxies in the photometric redshift catalog of DESI extragalactic sources \citep{Zou2019ApJS} that are located
within the FRB localization region and have r-band magnitude $>$ 22 AB mag along with their estimated photometric redshifts. All three galaxies have 5$\sigma$ lower limit on the redshift $>$ z$\rm{_{max}}$. {Therefore, we conclude that the association between FRB and NGC 3252 is real and robust.}
\begin{table}[ht]
\caption{Galaxies with m$_{\rm r} >$ 22 AB mag in the FRB 90\% confidence localization region.}
\label{tab:faint-galaxies}
\begin{center}
\hspace{-1.in}
\begin{tabular}{@{} *6c @{}}
\toprule
\textbf{Number} & \textbf{R.A.} &\textbf{Dec.}& \textbf{DESI(r-band)}&\textbf{z$_{\mathrm{photoz}}^a$}&\textbf{z$_{\mathrm{photoz-err}}^a,b$}\\
&J2000& J2000& AB mag.& &\\ \midrule 

1 & \rm{10$^{h}$34$^{m}$16$^{s}$.01} & 73$^{\degrees}$44$^{\arcmin}$18$^{\arcsec}$.60 & 22.56 & 0.62 & 0.06\\
2 & \rm{10$^{h}$33$^{m}$59$^{s}$.30} & 73$^{\degrees}$44$^{\arcmin}$40$^{\arcsec}$.56 & 22.90 & 0.60 & 0.09\\
3 & \rm{10$^{h}$34$^{m}$35$^{s}$.06} & 73$^{\degrees}$44$^{\arcmin}$58$^{\arcsec}$.56 & 22.73 & 0.72 & 0.06\\
\bottomrule 
 \hline
\end{tabular}
\end{center}
$^a$ From photometric redshift catalog of galaxies detected in the DESI survey \citep{Zou2019ApJS}.\\
$^b$ For all three galaxies, z$_{\mathrm{photz}} - 5\sigma_{\mathrm{photz-err}} > $ z$_{\rm max}$.

\end{table}
  

\subsection{Physical properties of NGC 3252}
\label{sec:physical-properties}

Here we summarise 
major physical properties of NGC 3252. 
We obtained long-slit  spectroscopy data from GTC, and its analysis is described in Appendix \ref{sec:longslit}. 
From the integrated optical spectrum of the galaxy, we estimate the oxygen abundance 12+$\log$(O/H)~=~8.44$\pm$0.06 (or nebular metallicity $\log(Z_{\rm gas}/Z_\odot)=-0.25\pm0.07$), which is $\sim$60\% of the solar value \citep{Asplund2009}. We also derive dust extinction at the V-band, A$_{v} =  1.3 \pm 0.2$ (E(B$-$V)$=0.42\pm0.06$ using R$_{v}=3.1$), using H$\alpha$/H$\beta$ ratio (i.e. Balmer decrement), assuming the standard Milky Way extinction curve \citep{Cardelli1989}. Finally, using SFR(H$\alpha$) = 7.9$\times10^{-42}$ M$_{\odot}$\,yr$^{-1} \times$ L(H$\alpha$/erg s$^{-1}$) \citep{kennicutt1994ApJ}, we get SFR(H$\alpha$)=0.033~M$_{\odot}$\,yr$^{-1}$ using the extinction corrected total H$\alpha$ luminosity of L(H$\alpha$) = 4.12$\times 10^{39}$~erg s$^{-1}$.


However, as the slits only cover a small fraction of the surface area of NGC 3252, it is expected that the above star formation rate is significantly underestimated. Therefore, we estimate the total star formation rate (SFR$\rm{_{total}}$) by combining the total infrared (TIR) luminosity and far-UV (FUV)-derived SFR as described in \cite{Iglesias2006}, which is found to be a robust estimate for the disk galaxies \citep{buat2007}. 
We estimated the TIR luminosity of NGC 3252 using the prescription of \cite{dale2002} which uses the Infrared Astronomical Satellite (IRAS) filters' fluxes \citep{Fullmer1995}, and got L(TIR) = 2.13 $\times 10^{9}$ L$_{\odot}$. Using Equation 5 from \cite{Iglesias2006}, SFR(TIR) = 0.38 M$_{\odot}$ yr$^{-1}$. Similarly, for the FUV luminosity, we use the Galex NUV filter flux and estimated the SFR(FUV) = 0.13 M$_{\odot}$ yr$^{-1}$  using the extinction uncorrected L(FUV) of NGC 3252 = 2.7 $\times$ 10$^{8}$  L$_{\odot}$. Finally, the total recent star formation rate was calculated using the relation from \cite{Iglesias2006}: SFR$_{total}$ = SFR(NUV) + (1-$\eta$) $\times$ SFR(TIR) = 0.36 M$_{\odot}$ yr$^{-1}$ where $\eta =$ 0.4 for disk galaxies in the local Universe \citep{Iglesias2004A&A}, which accounts for the fraction of the total IR luminosity heated by old stars. This relation has a calibrated uncertainty of about 20\%.    

\begin{figure}[ht]
\begin{center}
    \includegraphics[width=0.95\textwidth]{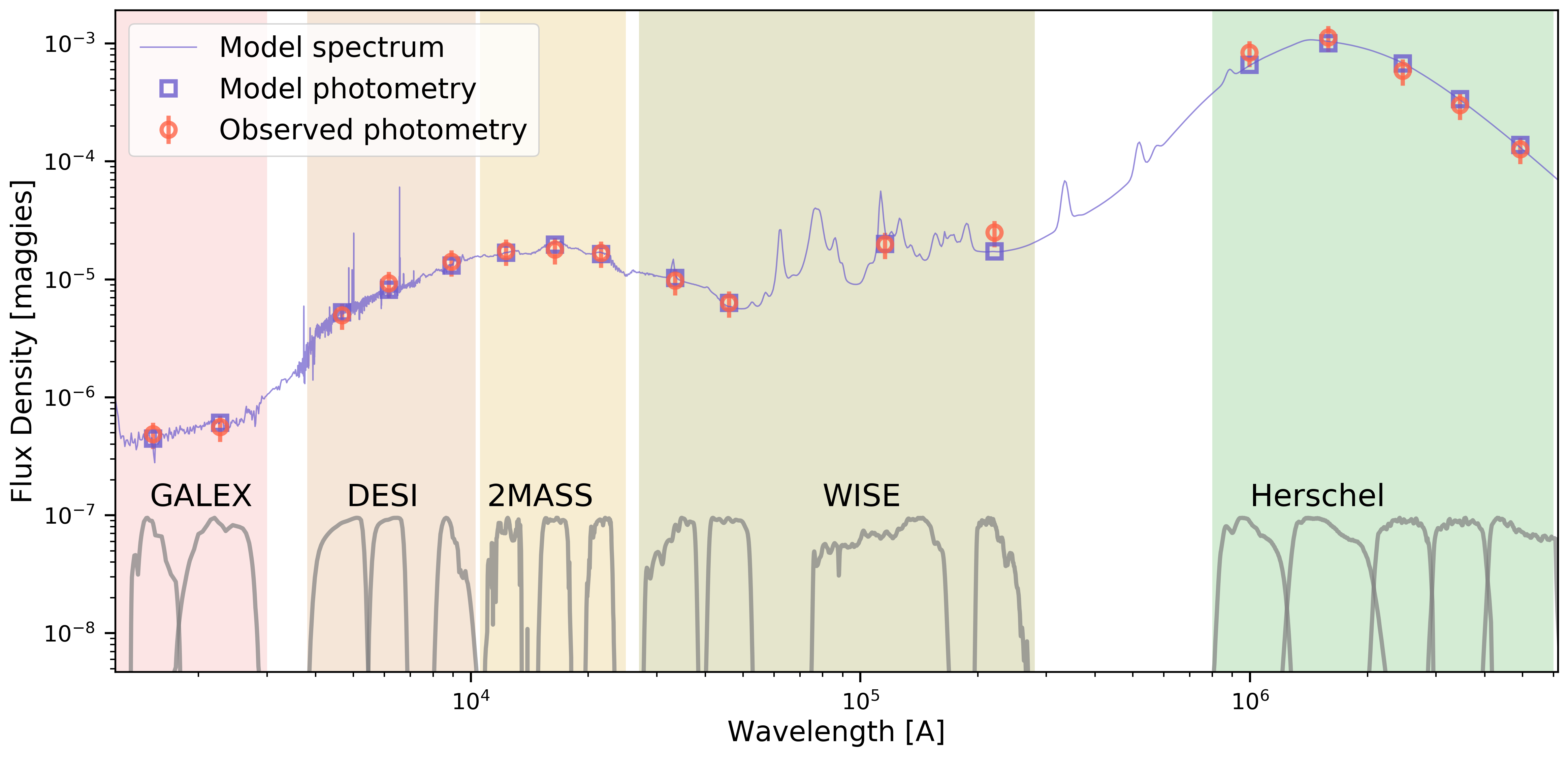}    
\end{center}
\caption{Modelling the SED of NGC 3252. The flux density 
of NGC 3252 in different wavelength bands are plotted along with the best-fit {\tt Prospector} model spectrum. To assess the quality of the {\tt Prospector} model, the modelled and actual photometry data are also shown. The shown model profile is used to estimate different physical properties of NGC 3252. For more information, see Appendix \ref{app:prospector}. The modelled SED of NGC 3252 shown in Figure \ref{fig:sed}
is in excellent agreement with that of a typical star-forming galaxy \citep{leitherer2005}.}
\label{fig:sed}
\end{figure}

To estimate stellar mass, metallicity and mass-weighted age of NGC 3252, we use a Bayesian inference spectral energy distribution (SED) fitting code, \texttt{Prospector} \citep{Leja2017,prospect2019}. Appendix \ref{app:prospector} describes the SED fitting analysis in detail.
We used 17 broadband filter fluxes covering the far-ultraviolet band (FUV) to far-infrared (FIR) bands (see Table \ref{sed:maggies}) of NGC 3252 and fit a five-parameter delayed-$\tau$ model \citep{simha2014,carnall2019ApJ}. This model and assumed priors of the free-parameters are discussed in Appendix \ref{app:prospector}. The best-fit spectral energy distribution (SED) profile of NGC 3252 is shown in Figure~\ref{fig:sed}. {\tt Prospector} also allows for Markov Chain Monte Carlo (MCMC) posterior sampling to estimate uncertainty in the best-fit values of the model parameters, which are stated in Table.~\ref{tab:host-properties}.



\begin{table}[ht]
\begin{center}
\caption{Notable properties of NGC 3252.}
\label{tab:host-properties}
\begin{tabular}{@{} *3l @{}}\toprule
\textbf{Property} & \textbf{Value} & \textbf{Reference}\\\midrule
log[SFR] ($\textup{M}_\odot\ \mathrm{yr^{-1}}$)$^{a}$ & $-0.45$ $\pm$ 0.1 & this work \\ 
Stellar Metallicity ($\mathrm{log}(\text{Z}/\text{Z}_{\odot})$)$^{a}$ & $-0.21^{+0.18}_{-0.19}$ &  this work\\
Nebular Metallicity ($\mathrm{\log(Z_{\rm gas}/Z_\odot)}$) & $-0.25\pm$ 0.07 & this work\\

Oxygen abundance [O/H] & 8.44 $\pm$ 0.06 & this work\\
Stellar mass ($\rm{M}_{\odot})$ & $ 5.8^{+1.6}_{-2.0} \times 10^{9}$  & this work\\
Effective radius (R$_{\mathrm{eff}}$; kpc) &  2.6 & \cite{salo2015}\\
Mass-weighted age (Gyr)$^{a}$ &  4.8$^{+1.6}_{-1.8}$  & this work\\

E(B-V) (mag) & 0.42 $\pm$ 0.06 & this work \\

Absolute r-band mag. (AB) & $-$19.1 $\pm$ 0.5 & -- \\
Luminosity distance (Mpc) & 20 $\pm$ 5 & \cite{Tully2016} \\\bottomrule 
 \hline
\end{tabular}
\end{center}
$^{a}$ Estimated using \texttt{Prospector}; See Appendix \ref{app:prospector}.
\end{table}

\subsection{Search for a multi-wavelength counterpart to FRB 20181030A}
\label{sec:multi-wavelength}

\subsubsection{Persistent radio source search}
\label{subsection:radio}
We searched archival radio data of the following surveys to check for the presence of a persistent radio source within the FRB 
uncertainty region: the NRAO VLA Sky Survey  \citep[NVSS;][]{condon1998nrao}, the VLA Sky Survey  \citep[VLASS;]
{lacy2016vla}, the Westerbork Northern Sky Survey \citep[WENSS;][]{rengelink1997westerbork}, and the Tata Institute of Fundamental Research Giant Metrewave Radio Telescope Sky Survey (TGSS) Alternative Data Release \citep[][]{intema2017gmrt}. We found only one radio source, NVSS J103422+734554. The radio source is only detected in NVSS and is either unresolved or marginally extended. Moreover, it is spatially coincident with the center of NGC 3252. Table \ref{tab:radio-data} 
lists 5$\sigma$ upper limits on the source's integrated flux density derived from the archival radio images of all other surveys. The NVSS radio source is likely resolved out and hence, undetected in the VLASS 2.1 data. In VLASS 1.1 data, we detected an irregular-shape source spatially coincident with the NVSS radio source. However, due to the lack of detection in the VLASS 2.1 data despite similar sensitivity, and known calibration and imaging artefacts in the VLASS 1.1 data \citep{lacy2016vla}, the radio source is likely spurious (M. Lacy, private communication). From the non-detection in the TGSS data and assuming a power-law dependence of the NVSS radio source flux density i.e., S$_{\nu} \propto \mathrm{S}^{\alpha}$, we estimated a lower limit on $\alpha > -0.43$. This agrees well with the observed radio continuum spectral index of local star-forming galaxies \citep[between $-$0.1 and $-$0.7;][]{Marvil2015AJ}. 

\begin{figure}[ht]
\centering
\includegraphics[width=.75\linewidth]{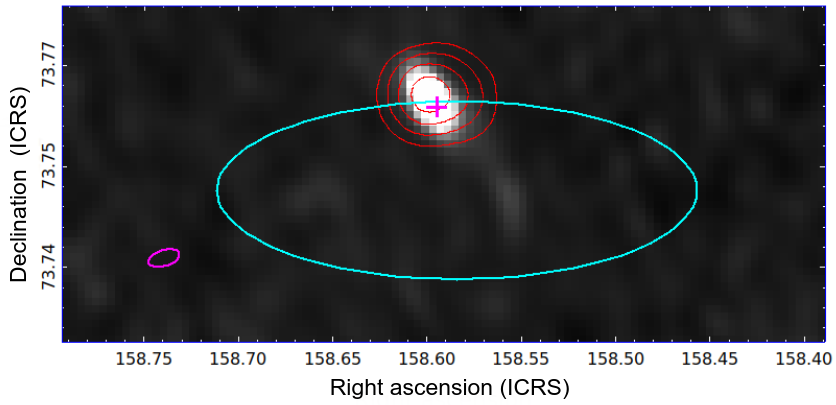}
\caption{The EVLA 4.5 GHz image of the FRB 20181030A 90\% localization region (cyan ellipse). NVSS contours (3 mJy, 2.5 mJy, 2 mJy, and 1.5 mJy) of the radio source are shown in red. The center of NGC 3252 (see Table \ref{Tab2}) is represented by a magenta cross. Finally, the EVLA beam is shown as magenta ellipse on the bottom left side of the image.
}
\label{fig:radio_image}
\end{figure}

While searching the VLA archive, we also found raw EVLA data (project ID = AK752) that cover the FRB localization region. Observations were conducted on 2010 June 19 (MJD 55366) with the array in D-configuration in two 128-MHz bandwidth sub-bands with central frequencies 4.495 GHz and 7.852 GHz, and about 40 minutes of time on source. The absolute flux density calibrator 3C147 and the phase calibrator J1048+7143 were used. The data were calibrated and flagged using \texttt{CASA} software \citep{casa2007}. Additional RFI flagging and self-calibration were done resulting in a final primary-beam corrected image with a local rms noise of $\sigma \approx$ 30 $\mu$Jy beam$^{-1}$.
Within the FRB localization region, we detect only the NVSS radio source extended in both the EVLA observations (See Figure \ref{fig:radio_image}). The integrated flux density of the NVSS source at 4.495 GHz and 7.852 GHz is estimated using the \texttt{Aegean} package \citep{hancock2012compact,hancock2018source} and is stated in Table \ref{tab:radio-data}. Using the EVLA flux densities at 4.495 GHz and 7.852 GHz, we estimated $\alpha$ to be $-0.94 \pm 0.16$, which is steeper than the lower-limit on $\alpha$ estimated using the flux densities at 150 MHz and 1.4 GHz ($> -0.43$). This is not unusual as the radio spectra of star-forming galaxies are known to show a break (or an exponential decline) in the frequency range of 1$-$12 GHz \citep{2018A&A...611A..55K}. 

The NVSS source is likely to be produced via ongoing star formation in NGC 3252.  To test this, we estimate the SFR using the NVSS 1.4 GHz continuum emission and compare it with the value estimated in \S\ref{sec:longslit}. Using the 1.4 GHz–SFR relation from \cite{davis2017MNRAS}, log(SFR$_{\rm UV+TIR}/\rm{M_{\odot} yr^{-1}}) = \rm 0.66 \pm 0.02 \times log(L_{1.4}(W/Hz)) - 14.02 \pm 0.39$, we estimate log(SFR$_{\rm UV+TIR}/\rm{M_{\odot} yr^{-1}}) = -0.6 \pm 0.5$ which  
agrees with the SFR estimate in Table \ref{tab:host-properties}. 
Though it is difficult to rule out the presence of a low-luminosity active galactic nucleus (AGN) at the center of NGC 3252 \citep{2007MNRAS.377.1696M}, the extended nature of the radio source and agreement of its 1.4 GHz flux density with the SFR of NGC 3252 suggest that an AGN is unlikely to be the dominant source of the observed persistent radio emission.
Moreover, from the non-detection of a persistent compact radio source ($\rm{< 0.3~kpc}$ at 20 Mpc) in the FRB 20181030A localization region in the VLASS 2.1 data (which has the best angular resolution among all the radio surveys considered here), we estimate a 3$\sigma$ upper limit of 480 $\mu$Jy at 3 GHz which at 20 Mpc implies an isotropic spectral luminosity $\approx 2 \times 10^{26}$ erg/s/Hz, at least 1500 times fainter than that the persistent radio source detected spatially coincident to FRB 20121102A \citep{chatterjee2017direct,resmi2020}.

\begin{table}[ht]
\hspace{-1.in}

\caption{Summary of radio observations of NGC 3252.}
\label{tab:radio-data}
\begin{tabular}{ccccc}
\toprule
Survey & Frequency & Date & Image Resolution$^{a}$ & Integrated Flux Density \\ 
& GHz & UT & $\arcsec$ & mJy \\\midrule 
TGSS & 0.15 & 2016 March 15 & 25 & $<$10$^{b}$ \\
WENSS & 0.326 & 1997 October 22 & 56  &$<18^{b}$  \\ 
NVSS$^{c}$ & 1.4 & 1993 December 18 & 45 & $3.8 \pm 0.5$ \\
VLASS 2.1 & 3.0 & 2020 October 13 & 2.5 &$<0.6^{b}$ \\
EVLA$^{c}$ & 4.495 & 2010 June 19 & 12.8 & $1.35 \pm 0.06$ \\ 
& 7.852 & 2010 June 19 & 9.1 &$0.80 \pm 0.06$ \\ 
\bottomrule 
\end{tabular}

$^a$ For each survey, average of major and minor axes of the formed beam is quoted.\\
$^b$ 5$\times$ local rms noise.\\
$^c$ The lone radio source in the FRB uncertainty region is extended and spatially coincident with the center of NGC 3252 in the NVSS and two EVLA observations.\\

\end{table}

\subsubsection{Archival search for X-ray counterparts}
\label{section:X-ray-followups}
We searched the Transient Name Server (TNS)\footnote{\url{https://www.wis-tns.org/}} for any archival transient event that is spatially and temporally coincident with any of the nine recorded FRB 20181030A bursts and found none. 
We also checked if the FRB was visible to the {\it Swift}/Burst Alert Telescope (BAT) and {\it Fermi}/GBM  at the time of the bursts. Unfortunately,
{\it Swift}/BAT was either not operational (transiting through the South Atlantic Anomaly region) or the FRB location was not within the BAT's field-of-view. Similarly, for all except FRB 20200122A
, the FRB location was not visible to {\it Fermi}/GBM. If FRB 20200122A was associated with a giant magnetar flare like the one detected from SGR 1806$-$20 on 2004 December 27 \citep{palmer2005giant}, {\it Fermi}/GBM with a flux sensitivity of $\sim$ 10$^{-7}$ erg cm$^{-2}$ s$^{-1}$ in the 50–300 keV band \citep{2020ApJ...893...46V} would have marginally 
detected it.
This places an upper limit on the coincident X-ray flare energy $\approx$ 10$^{46}$ erg s$^{-1}$ at 20 Mpc (without correcting for the attenuation by the host). 
There is an X-ray source RX J1034.3.3+7345 \citep[or 1AXG J103422+7344 in the {\it ASCA} medium sensitivity survey by ][]{Ueda2001ApJS} in the vicinity of NGC 3252. This source was discovered in the {\it ROSAT} all-sky survey \citep{truemper1982rosat} and was initially associated with NGC 3252 by \cite{bade1998hamburg}, \cite{condom1998AJ} and \cite{bauer2000ApJS}. However, with the availability of higher resolution X-ray images, this association has been argued to be incorrect \citep{Haakonsen2009ApJS}. More interesting, the X-ray source was found to be spatially coincident with an optical transient PTF10hjz discovered by the Palomar Transient Factory (PTF) on 2010 May 16 \citep{ptf10hjz}. Based on its high optical and radio flux variability and estimated broadband spectral energy distribution, PTF10hjz was later classified as a background blazar (S. Kulkarni, private communication). Therefore, we conclude that the X-ray source RX J103423.1+734525 (or PTF10hjz) is unrelated to FRB 20181030A. 

\section{Discussion}
\label{sec:discussion}

\subsection{Constraints on the Milky Way halo DM contribution}

With NGC 3252 as its host, we can use FRB 20181030A and its low DM-excess \citep[as for FRB 20200120E;][]{bhardwaj2021}, to constrain the Milky Way halo DM along the FRB sight-line. 
At 20 Mpc, using the average Macquart relation, we estimate DM$_{\mathrm{IGM}} \approx$ 5 pc cm$^{-3}$.\footnote{DM$_{\mathrm{IGM}}$ is expected to be considerable as the FRB sight-line intersects several foreground groups, including that of M81 \citep{tully2015AJ} making DM$_{\mathrm{IGM}}$ = 5 pc cm$^{-3}$ a conservative estimate.} 
Assuming negligible host DM contribution, we find an upper limit on the DM$_{\mathrm{MW,halo}}$ to be 58 and 66 pc cm$^{-3}$ using the DM$_{\rm MW}$ estimate from the NE2001 and YMW16 models, respectively (See Table \ref{tab:params}). However, a negligible host contribution is likely an overly conservative assumption as even in the extreme scenario where the FRB has a very large offset from the host, the host's circumgalactic medium would still contribute to the FRB DM.
Therefore, we use the MCMC analysis discussed in Appendix \ref{app:mcmc}, but this time, fix the redshift of the FRB to that of NGC 3252. From this analysis, we estimate the DM$_{\rm MW,halo}$ 90\% Bayesian credible interval to be (19 pc cm$^{-3}$, 55 pc cm$^{-3}$).
This, along with a similar constraint derived by \cite{bhardwaj2021}, suggests that the Milky Way halo DM contribution could be relatively small.
This in turn would help in constraining the state and composition of the Milky Way circumgalactic medium \citep{tumlinson2017}. 
However, to constrain the average DM$_{\mathrm{MW,halo}}$ estimate, we need more low-DM FRBs.    

\subsection{Comparison with SGR 1935+2154 radio bursts}

From Table 2 of \cite{abb+19c}, the peak 400--800-MHz flux densities of the two published bursts from this source, FRBs 20181030A and 20181030B, are 3.2 $\pm$ 1.7 Jy and 3.1 $\pm$ 1.4 Jy, respectively.  At a distance of 20 Mpc, the isotropic radio luminosity of these two bursts would be $\sim 9\times $10$^{38}$ erg~s$^{-1}$, around six times larger than those of the very bright SGR 1935+2154 radio bursts recently detected by CHIME/FRB and STARE2 \citep{2020SGR,Bochenek2020}.\footnote{Assuming the distance to SGR 1935+2154 is 10 kpc, but note that \cite{Zhou2020ApJ}, \cite{mereghetti2020integral} and \cite{bailes2021} argue for a significantly smaller distance to the magnetar, $\approx 2 $-$ 7$ kpc.}
This suggests a continuum of FRB luminosities, at least at low values. \cite{Bochenek2020} estimated the volumetric rate of SGR 1935+2154-like bursts to be $7^{+9}_{-6} \times 10^{7} \; \mathrm{Gpc}^{-3} \mathrm{yr}^{-1}$, assuming that the FRB luminosity function follows a power law and the FRB rate is proportional to the star-formation rate. As at least the two FRB bursts 
in the first CHIME/FRB catalog \citep{firstchimefrbcatalog2021} have isotropic luminosity $\geq 10^{38}$ erg~s$^{-1}$, we estimate a lower limit on the volumetric rate of FRBs ($\geq 10^{38}$ erg~s$^{-1}$) to be $1.5^{+1.6}_{-0.7} \times 10^{7} \; \mathrm{Gpc}^{-3} \mathrm{yr}^{-1}$.
This lower limit is in agreement with the estimate by \cite{Bochenek2020}, which supports their conclusion that magnetars like those observed in the Milky Way could be a dominant channel of FRB production, at least at the lower end of the FRB luminosity function. 

Moreover, the estimated CHIME/FRB volumetric rate agrees with the rate calculated by extrapolating the luminosity function derived from a sample of bright FRBs observed at 1.4 GHz by the Australian Square Kilometre Array Pathfinder (ASKAP) and Parkes down to the luminosity of FRB 20181030A's bursts \citep{lu2019,luo2020}.  
Lastly, the estimated FRB volumetric rate at low luminosities is at least 100 times 
higher than the observed volumetric rate of core-collapse supernovae in the local Universe \citep[$\sim 10^{5}$~Gpc$^{-3}$~yr$^{-1}$]{taylor2014}. Assuming core-collapse supernovae are the most common way to produce compact objects, 
FRBs detected at low luminosities ($\sim 10^{38}$ erg~s$^{-1}$) are therefore more likely to be repeating sources.    
 
\subsection{Implications for different progenitor models}

Three repeating FRB sources within a comoving volume out to a distance of 20 Mpc \citep{2020SGR,Bochenek2020,bhardwaj2021} have now been discovered.
Using these discoveries, we estimate a lower limit on the comoving number density (n$_{\rm FRB}$) of repeating FRB sources to be 9$^{+7}_{-4} \times 10^{4}$ Gpc$^{-3}$. 
We can also express n$_{\rm FRB}$ 
= R$_{\rm FRB}\tau \eta \zeta$, where R$_{\rm FRB}$ is the local Universe volumetric birth rate of repeating FRB sources, $\tau$ and $\zeta$ are the average lifetime and active duty cycle of repeating FRBs, respectively, and $\eta$ is the beaming fraction. Taking the fiducial values of $\eta$ = 0.1 and $\zeta$ = 0.3 from \cite{lu2016MNRAS} and \cite{nicholl2017ApJ}, we estimate $\rm R_{ FRB}\tau = 3^{+2}_{-1} \times 10^{6}~ \Big(\frac{0.1}{\eta}\Big) \times \Big(\frac{0.3}{\zeta}\Big)~ Gpc^{-3}$. One of the popular proposed repeating FRB models is a highly magnetized ($> 10^{15}$ G) young neutron star with period $\sim$ ms. 
\cite{nicholl2017ApJ} estimated the volumetric birth-rate of millisecond magnetars (R$_{\rm ms}$) to be $\rm{ \sim few~10-100~Gpc^{-3} yr^{-1}}$.
Using this R$_{\rm ms}$ value as $\rm R_{ FRB}$, we estimate $\tau$ $\sim 10^{4} - 10^{5}$ yr. This is around two orders of magnitude greater than the expected typical lifetime of a repeating FRB in the models that invoke millisecond magnetars \citep[$\sim 30-300$ yrs;][]{metzger2017ApJ,metzger2019fast}. Therefore, it is unlikely that all repeating FRBs are produced by millisecond magnetars formed primarily via cataclysmic events, like superluminous supernovae, or long and short gamma-ray bursts. Note that \cite{nicholl2017ApJ} did not include accretion induced collapse (AIC) as a channel for forming millisecond magnetars in their calculation due to the high uncertainty in the AIC rates. 
However, theoretically estimated rates of AIC are found to be comparable to that of binary neutron star mergers \citep{Tauris2013A&A,Kwiatkowski2015}. If these estimates are correct, including them would not change our conclusion significantly.   

\subsection{Comparison with other repeating FRB hosts}

With the inclusion of FRB 20181030A, likely localized to the star-forming galaxy NGC 3252, in the sample of five repeating FRBs, 20121102A, 20180916B, 20190711A, 20200120E, and FRB 20201124A \citep{2021ATel14497....1C,2021arXiv210611993F,2021arXiv210609710R}, it is evident that the repeating FRB hosts exhibit a continuum of properties in terms of their luminosities, stellar masses, metallicity, and SFRs, ranging from FRB 20121102A, a metal-poor, high-star forming dwarf irregular galaxy, to FRB 20200120E, a metal-rich massive early-type spiral galaxy.
However, it is interesting to note that all five localized repeating FRBs discovered thus far are in either spiral or irregular galaxies \citep{Mannings+21,bhardwaj2021}, where practically all core-collapse supernovae (SN II, IIn, IIb, and Ib/c) occur \citep{Bergh2005PASP}.
However, FRB 20200120E is localized to an M81 globular cluster \citep{Kristan2021arXiv} where core-collapse supernovae are not expected to occur. Therefore, we need a larger sample of FRB hosts to decipher the nature of FRB progenitors.
Lastly, we note that all three local Universe repeating FRBs 
have thus far been observed to produce only low-energy bursts ($\lesssim 10^{35}$ erg), unlike, for example, the FRB 20121102A bursts, which have shown a range of burst energies \citep[10$^{36}-10^{40}$ erg; e.g.,][]{chatterjee2005getting, gourdji2019ApJ}. More bursts, particularly high-energy ones, from these FRBs would aid in constraining the emission mechanism of the local Universe FRBs \citep{2021yuri}. 

\section{Summary \& Conclusions}
\label{sec:conclusion}

We have reported on the likely association of the repeating FRB 20181030A discovered by \cite{abb+19c} with a nearby star-forming spiral galaxy, NGC 3252, at a distance of 20 Mpc. The chance coincidence probability of finding NGC 3252 within the FRB localization region is $< 2.5 \times 10^{-3}$. Moreover, we searched for plausible host galaxies within the 90\% confidence localization region of the FRB, and found no galaxy except NGC 3252 with M$_{\rm r} < -15$, a limit in luminosity over five times smaller than for any FRB hosts identified to date. 

NGC 3252 is a star-forming spiral galaxy (see Figure \ref{fig:sed}). 
We found no archival transient event spatially or temporally coincident with any of the reported FRB 20181030A bursts to date. For one FRB burst, FRB 20200122A, that was detected on 2020 January 22 by CHIME/FRB and was also visible to {\it Fermi}/GBM, we estimated an upper limit on the coincident X-ray flare energy to be $\approx$ 10$^{46}$ erg s$^{-1}$. We also searched for a compact, persistent radio continuum source within the FRB localization region and found none. We then estimated a 3$\sigma$ upper limit at 3 GHz $\rm= 4.3 \times 10^{25}~erg~s^{-1} Hz^{-1}$, at least 1500 times fainter than the persistent source associated with FRB 20121102A. Due to its low DM-excess, we constrain the Milky Way halo DM contribution to be 19$-$55 pc cm$^{-3}$ (90\% confidence interval) along the FRB sight-line.  We also compared the two published FRB 20181030A bursts with those of SGR 1935+2154. The FRB bursts' isotropic luminosity is $\sim$ 6 times larger than those of SGR 1935+2154; using this, we have estimated a lower limit on the volumetric rate of FRBs with luminosities $\geq 10^{38}$ erg s$^{-1}$. We found this to be in good agreement with the rate estimated by \cite{Bochenek2020} using the SGR 1935+2154 radio burst, suggesting that many low-luminosity FRBs could be produced by magnetars. Lastly, we also showed that it is unlikely that most of the repeating FRB progenitors are young millisecond magnetars, and that if we expect millisecond magnetars to be a source of repeating FRBs, we need multiple repeating FRB formation channels.

At a distance of 20 Mpc, FRB 20181030A is one of the closest FRBs discovered to date. In principle, it should be possible to detect prompt multi-wavelength counterparts as predicted by several FRB models \citep{yi2014multi,burke2018multiple,chen2020multiwavelength,nicastro2021}. Therefore, we strongly encourage multi-wavelength follow-up of FRB 20181030A.

\facility{CHIME, Fermi, GTC, VLA}

\software{ Prospector \citep{Leja2017,prospect2019}, FSPS \citep{Conroy2009}, MARZ \citep{Hinton}, CASA \citep{casa2007}, Aegean \citep{hancock2012compact,hancock2018source}, emcee \citep{fm2013}, IRAF \citep{tody1986iraf,tody1993iraf}, Fermi GBM data tools \citep{GbmDataTools}, Astropy \citep{astropy:2013,astropy:2018}, ALPpy \citep{2012ascl.soft08017R}, SAOImage DS9 \citep{2003ASPC..295..489J}, NumPy \citep{harris2020array}, Matplotlib \citep{Hunter:2007}}

\acknowledgements
We thank the anonymous reviewer and ApJL data editor for their careful reading of our manuscript and their insightful comments and suggestions.
We also express our gratitude to  Dr. Mark Lacy for the useful discussion related to the VLASS 1.1 data.
We thank the Dominion Radio Astrophysical Observatory, operated by the National
Research Council Canada, for gracious hospitality and expertise. We acknowledge that CHIME is located on the traditional, ancestral, and unceded territory of the Syilx/Okanagan people. CHIME is funded by a grant from the Canada Foundation for Innovation (CFI) 2012 Leading Edge Fund (Project 31170) and by contributions from the provinces of British Columbia, Qu\'ebec and Ontario. The CHIME/FRB Project is funded by a grant from the CFI 2015 Innovation Fund (Project 33213) and by contributions from the provinces of British Columbia and Qu\'ebec, and by the Dunlap Institute for Astronomy and Astrophysics at the University of Toronto. Additional support was provided by the Canadian Institute for Advanced Research (CIFAR), McGill University and the McGill Space Institute via the Trottier Family Foundation, and the University of British Columbia. The Dunlap Institute is funded through an endowment established by the David Dunlap family and the University of Toronto. Research at Perimeter Institute is supported by the Government of Canada through Industry Canada and by the Province of Ontario through the Ministry of Research \& Innovation.  FRB research at UBC is supported by an NSERC Discovery Grant and by the Canadian Institute for Advanced Research.  The CHIME/FRB baseband system is funded in part by a Canada Foundation for Innovation John R. Evans Leaders Fund award to I.H.S. The National Radio Astronomy Observatory is a facility of the National Science Foundation (NSF) operated under cooperative agreement by Associated Universities, Inc.  
Based on observations made with the GTC telescope, 
in the Spanish Observatorio del Roque de los Muchachos of the Instituto de Astrofísica de Canarias, under Director’s Discretionary Time.
Based on observations made with the Gran Telescopio Canarias (GTC), installed at the Spanish Observatorio del Roque de los Muchachos of the Instituto de Astrofísica de Canarias, in the island of La Palma.
This research has made use of the NASA/IPAC Extragalactic Database (NED), which is operated by the Jet Propulsion Laboratory, California Institute of Technology, under contract with the National Aeronautics and Space Administration. 

M.B. is supported by an FRQNT Doctoral Research Award. D.M. is a Banting Fellow. T.D.M is grateful to CONACYT  for  the  research  grant  CB-A1-S-25070. V.M.K. holds the Lorne Trottier Chair in Astrophysics \& Cosmology, a Distinguished James McGill Professorship and receives support from an NSERC Discovery Grant (RGPIN 228738-13) and Gerhard Herzberg Award, from an R. Howard Webster Foundation Fellowship from CIFAR, and from the FRQNT CRAQ.
and from the FRQNT Centre de Recherche en Astrophysique du Quebec. B.M.G. is supported by an NSERC Discovery Grant (RGPIN-2015-05948), and by the Canada Research Chairs (CRC) program. C.L. was supported by the U.S. Department of Defense (DoD) through the National Defense Science $\&$ Engineering Graduate Fellowship (NDSEG) Program. M.M. is supported by an NSERC PGS-D award. E.P. acknowledges funding from an NWO Veni Fellowship. P.S. is a Dunlap Fellow and an NSERC Postdoctoral Fellow. K.S. is supported by the NSF Graduate Research Fellowship Program. 


\bibliographystyle{aasjournal}
\bibliography{ref.bib}

\begin{thebibliography}{}
\expandafter\ifx\csname natexlab\endcsname\relax\def\natexlab#1{#1}\fi
\providecommand{\url}[1]{\href{#1}{#1}}

\bibitem[{{Arora} {et~al.}(2021){Arora}, {Stone}, {Courteau}, \&
  {Jarrett}}]{arora2021}
{Arora}, N., {Stone}, C., {Courteau}, S., \& {Jarrett}, T.~H. 2021, arXiv
  e-prints, arXiv:2105.01660

\bibitem[{{Asplund} {et~al.}(2009){Asplund}, {Grevesse}, {Sauval}, \&
  {Scott}}]{Asplund2009}
{Asplund}, M., {Grevesse}, N., {Sauval}, A.~J., \& {Scott}, P. 2009, \araa, 47,
  481

\bibitem[{{Astropy Collaboration} {et~al.}(2013){Astropy Collaboration},
  {Robitaille}, {Tollerud}, {Greenfield}, {Droettboom}, {Bray}, {Aldcroft},
  {Davis}, {Ginsburg}, {Price-Whelan}, {Kerzendorf}, {Conley}, {Crighton},
  {Barbary}, {Muna}, {Ferguson}, {Grollier}, {Parikh}, {Nair}, {Unther},
  {Deil}, {Woillez}, {Conseil}, {Kramer}, {Turner}, {Singer}, {Fox}, {Weaver},
  {Zabalza}, {Edwards}, {Azalee Bostroem}, {Burke}, {Casey}, {Crawford},
  {Dencheva}, {Ely}, {Jenness}, {Labrie}, {Lim}, {Pierfederici}, {Pontzen},
  {Ptak}, {Refsdal}, {Servillat}, \& {Streicher}}]{astropy:2013}
{Astropy Collaboration}, {Robitaille}, T.~P., {Tollerud}, E.~J., {et~al.} 2013,
  \aap, 558, A33

\bibitem[{{Astropy Collaboration} {et~al.}(2018){Astropy Collaboration},
  {Price-Whelan}, {Sip{\H{o}}cz}, {G{\"u}nther}, {Lim}, {Crawford}, {Conseil},
  {Shupe}, {Craig}, {Dencheva}, {Ginsburg}, {Vand erPlas}, {Bradley},
  {P{\'e}rez-Su{\'a}rez}, {de Val-Borro}, {Aldcroft}, {Cruz}, {Robitaille},
  {Tollerud}, {Ardelean}, {Babej}, {Bach}, {Bachetti}, {Bakanov}, {Bamford},
  {Barentsen}, {Barmby}, {Baumbach}, {Berry}, {Biscani}, {Boquien}, {Bostroem},
  {Bouma}, {Brammer}, {Bray}, {Breytenbach}, {Buddelmeijer}, {Burke},
  {Calderone}, {Cano Rodr{\'\i}guez}, {Cara}, {Cardoso}, {Cheedella}, {Copin},
  {Corrales}, {Crichton}, {D'Avella}, {Deil}, {Depagne}, {Dietrich}, {Donath},
  {Droettboom}, {Earl}, {Erben}, {Fabbro}, {Ferreira}, {Finethy}, {Fox},
  {Garrison}, {Gibbons}, {Goldstein}, {Gommers}, {Greco}, {Greenfield},
  {Groener}, {Grollier}, {Hagen}, {Hirst}, {Homeier}, {Horton}, {Hosseinzadeh},
  {Hu}, {Hunkeler}, {Ivezi{\'c}}, {Jain}, {Jenness}, {Kanarek}, {Kendrew},
  {Kern}, {Kerzendorf}, {Khvalko}, {King}, {Kirkby}, {Kulkarni}, {Kumar},
  {Lee}, {Lenz}, {Littlefair}, {Ma}, {Macleod}, {Mastropietro}, {McCully},
  {Montagnac}, {Morris}, {Mueller}, {Mumford}, {Muna}, {Murphy}, {Nelson},
  {Nguyen}, {Ninan}, {N{\"o}the}, {Ogaz}, {Oh}, {Parejko}, {Parley}, {Pascual},
  {Patil}, {Patil}, {Plunkett}, {Prochaska}, {Rastogi}, {Reddy Janga},
  {Sabater}, {Sakurikar}, {Seifert}, {Sherbert}, {Sherwood-Taylor}, {Shih},
  {Sick}, {Silbiger}, {Singanamalla}, {Singer}, {Sladen}, {Sooley},
  {Sornarajah}, {Streicher}, {Teuben}, {Thomas}, {Tremblay}, {Turner},
  {Terr{\'o}n}, {van Kerkwijk}, {de la Vega}, {Watkins}, {Weaver}, {Whitmore},
  {Woillez}, {Zabalza}, \& {Astropy Contributors}}]{astropy:2018}
{Astropy Collaboration}, {Price-Whelan}, A.~M., {Sip{\H{o}}cz}, B.~M., {et~al.}
  2018, \aj, 156, 123

\bibitem[{Bade {et~al.}(1998)Bade, Engels, Voges, Beckmann, Boller, Cordis,
  Dahlem, Englhauser, Molthagen, Nass, {et~al.}}]{bade1998hamburg}
Bade, N., Engels, D., Voges, W., {et~al.} 1998, Astronomy and Astrophysics
  Supplement Series, 127, 145

\bibitem[{{Bailes} {et~al.}(2021){Bailes}, {Bassa}, {Bernardi}, {Buchner},
  {Burgay}, {Caleb}, {Cooper}, {Desvignes}, {Groot}, {Heywood}, {Jankowski},
  {Karuppusamy}, {Kramer}, {Malenta}, {Naldi}, {Pilia}, {Pupillo}, {Rajwade},
  {Spitler}, {Surnis}, {Stappers}, {Addis}, {Bloemen}, {Bezuidenhout},
  {Bianchi}, {Champion}, {Chen}, {Driessen}, {Geyer}, {Gourdji}, {Hessels},
  {Kondratiev}, {Klein-Wolt}, {K{\"o}rding}, {Le Poole}, {Liu}, {Lower},
  {Lyne}, {Magro}, {McBride}, {Mickaliger}, {Morello}, {Parthasarathy},
  {Paterson}, {Perera}, {Pieterse}, {Pleunis}, {Possenti}, {Rowlinson},
  {Serylak}, {Setti}, {Tavani}, {Wijers}, {ter Veen}, {Venkatraman Krishnan},
  {Vreeswijk}, \& {Woudt}}]{bailes2021}
{Bailes}, M., {Bassa}, C.~G., {Bernardi}, G., {et~al.} 2021, \mnras, 503, 5367

\bibitem[{{Bauer} {et~al.}(2000){Bauer}, {Condon}, {Thuan}, \&
  {Broderick}}]{bauer2000ApJS}
{Bauer}, F.~E., {Condon}, J.~J., {Thuan}, T.~X., \& {Broderick}, J.~J. 2000,
  \apjs, 129, 547

\bibitem[{{Bhardwaj} {et~al.}(2021){Bhardwaj}, {Gaensler}, {Kaspi},
  {Landecker}, {Mckinven}, {Michilli}, {Pleunis}, {Tendulkar}, {Andersen},
  {Boyle}, {Cassanelli}, {Chawla}, {Cook}, {Dobbs}, {Fonseca}, {Kaczmarek},
  {Leung}, {Masui}, {Mnchmeyer}, {Ng}, {Rafiei-Ravandi}, {Scholz}, {Shin},
  {Smith}, {Stairs}, \& {Zwaniga}}]{bhardwaj2021}
{Bhardwaj}, M., {Gaensler}, B.~M., {Kaspi}, V.~M., {et~al.} 2021, \apjl, 910,
  L18

\bibitem[{{Bochenek} {et~al.}(2020){Bochenek}, {Ravi}, {Belov}, {Hallinan},
  {Kocz}, {Kulkarni}, \& {McKenna}}]{Bochenek2020}
{Bochenek}, C.~D., {Ravi}, V., {Belov}, K.~V., {et~al.} 2020, arXiv e-prints,
  arXiv:2005.10828

\bibitem[{{Bohlin} {et~al.}(1995){Bohlin}, {Colina}, \& {Finley}}]{std3}
{Bohlin}, R.~C., {Colina}, L., \& {Finley}, D.~S. 1995, \aj, 110, 1316

\bibitem[{{Buat} {et~al.}(2007){Buat}, {Takeuchi}, {Iglesias-P{\'a}ramo}, {Xu},
  {Burgarella}, {Boselli}, {Barlow}, {Bianchi}, {Donas}, {Forster}, {Friedman},
  {Heckman}, {Lee}, {Madore}, {Martin}, {Milliard}, {Morissey}, {Neff}, {Rich},
  {Schiminovich}, {Seibert}, {Small}, {Szalay}, {Welsh}, {Wyder}, \&
  {Yi}}]{buat2007}
{Buat}, V., {Takeuchi}, T.~T., {Iglesias-P{\'a}ramo}, J., {et~al.} 2007, \apjs,
  173, 404

\bibitem[{Burke-Spolaor(2018)}]{burke2018multiple}
Burke-Spolaor, S. 2018, Nature Astronomy, 2, 845

\bibitem[{{Byler} {et~al.}(2017){Byler}, {Dalcanton}, {Conroy}, \&
  {Johnson}}]{byler2017}
{Byler}, N., {Dalcanton}, J.~J., {Conroy}, C., \& {Johnson}, B.~D. 2017, \apj,
  840, 44

\bibitem[{{Calzetti} {et~al.}(2000){Calzetti}, {Armus}, {Bohlin}, {Kinney},
  {Koornneef}, \& {Storchi-Bergmann}}]{Calzetti2000ApJ}
{Calzetti}, D., {Armus}, L., {Bohlin}, R.~C., {et~al.} 2000, \apj, 533, 682

\bibitem[{{Cardelli} {et~al.}(1989){Cardelli}, {Clayton}, \&
  {Mathis}}]{Cardelli1989}
{Cardelli}, J.~A., {Clayton}, G.~C., \& {Mathis}, J.~S. 1989, \apj, 345, 245

\bibitem[{{Carnall} {et~al.}(2019){Carnall}, {Leja}, {Johnson}, {McLure},
  {Dunlop}, \& {Conroy}}]{carnall2019ApJ}
{Carnall}, A.~C., {Leja}, J., {Johnson}, B.~D., {et~al.} 2019, \apj, 873, 44

\bibitem[{Chatterjee {et~al.}(2005)Chatterjee, Vlemmings, Brisken, Lazio,
  Cordes, Goss, Thorsett, Fomalont, Lyne, \& Kramer}]{chatterjee2005getting}
Chatterjee, S., Vlemmings, W., Brisken, W., {et~al.} 2005, The Astrophysical
  Journal Letters, 630, L61

\bibitem[{Chatterjee {et~al.}(2017)Chatterjee, Law, Wharton, Burke-Spolaor,
  Hessels, Bower, Cordes, Tendulkar, Bassa, Demorest,
  {et~al.}}]{chatterjee2017direct}
Chatterjee, S., Law, C., Wharton, R., {et~al.} 2017, Nature, 541, 58

\bibitem[{Chen {et~al.}(2020)Chen, Ravi, \& Lu}]{chen2020multiwavelength}
Chen, G., Ravi, V., \& Lu, W. 2020, arXiv preprint arXiv:2004.10787

\bibitem[{{CHIME/FRB Collaboration}(2021)}]{2021ATel14497....1C}
{CHIME/FRB Collaboration}. 2021, The Astronomer's Telegram, 14497, 1

\bibitem[{{CHIME/FRB Collaboration} {et~al.}(2018){CHIME/FRB Collaboration},
  {Amiri}, {Bandura}, {Berger}, {Bhardwaj}, {Boyce}, {Boyle}, {Brar},
  {Burhanpurkar}, {Chawla}, {Chowdhury}, {Cliche}, {Cranmer}, {Cubranic},
  {Deng}, {Denman}, {Dobbs}, {Fandino}, {Fonseca}, {Gaensler}, {Giri},
  {Gilbert}, {Good}, {Guliani}, {Halpern}, {Hinshaw}, {H{\"o}fer}, {Josephy},
  {Kaspi}, {Landecker}, {Lang}, {Liao}, {Masui}, {Mena-Parra}, {Naidu},
  {Newburgh}, {Ng}, {Patel}, {Pen}, {Pinsonneault-Marotte}, {Pleunis}, {Rafiei
  Ravandi}, {Ransom}, {Renard}, {Scholz}, {Sigurdson}, {Siegel}, {Smith},
  {Stairs}, {Tendulkar}, {Vand erlinde}, \& {Wiebe}}]{abb+2018ApJ}
{CHIME/FRB Collaboration}, {Amiri}, M., {Bandura}, K., {et~al.} 2018, \apj,
  863, 48

\bibitem[{{CHIME/FRB Collaboration} {et~al.}(2019{\natexlab{a}}){CHIME/FRB
  Collaboration}, {Andersen}, {Band ura}, {Bhardwaj}, {Boubel}, {Boyce},
  {Boyle}, {Brar}, {Cassanelli}, {Chawla}, {Cubranic}, {Deng}, {Dobbs},
  {Fandino}, {Fonseca}, {Gaensler}, {Gilbert}, {Giri}, {Good}, {Halpern},
  {H{\"o}fer}, {Hill}, {Hinshaw}, {Josephy}, {Kaspi}, {Kothes}, {Landecker},
  {Lang}, {Li}, {Lin}, {Masui}, {Mena-Parra}, {Merryfield}, {Mckinven},
  {Michilli}, {Milutinovic}, {Naidu}, {Newburgh}, {Ng}, {Patel}, {Pen},
  {Pinsonneault-Marotte}, {Pleunis}, {Rafiei-Ravandi}, {Rahman}, {Ransom},
  {Renard}, {Scholz}, {Siegel}, {Singh}, {Smith}, {Stairs}, {Tendulkar},
  {Tretyakov}, {Vanderlinde}, {Yadav}, \& {Zwaniga}}]{abb+19c}
{CHIME/FRB Collaboration}, {Andersen}, B.~C., {Band ura}, K., {et~al.}
  2019{\natexlab{a}}, \apjl, 885, L24

\bibitem[{{CHIME/FRB Collaboration} {et~al.}(2019{\natexlab{b}}){CHIME/FRB
  Collaboration}, {Andersen}, {Bandura}, {Bhardwaj}, {Boubel}, {Boyce},
  {Boyle}, {Brar}, {Cassanelli}, {Chawla}, {Cubranic}, {Deng}, {Dobbs},
  {Fandino}, {Fonseca}, {Gaensler}, {Gilbert}, {Giri}, {Good}, {Halpern},
  {Hill}, {Hinshaw}, {H{\"o}fer}, {Josephy}, {Kaspi}, {Kothes}, {Landecker},
  {Lang}, {Li}, {Lin}, {Masui}, {Mena-Parra}, {Merryfield}, {Mckinven},
  {Michilli}, {Milutinovic}, {Naidu}, {Newburgh}, {Ng}, {Patel}, {Pen},
  {Pinsonneault-Marotte}, {Pleunis}, {Rafiei-Ravandi}, {Rahman}, {Ransom},
  {Renard}, {Scholz}, {Siegel}, {Singh}, {Smith}, {Stairs}, {Tendulkar},
  {Tretyakov}, {Vanderlinde}, {Yadav}, \& {Zwaniga}}]{andersen2019chime}
{CHIME/FRB Collaboration}, {Andersen}, B.~C., {Bandura}, K., {et~al.}
  2019{\natexlab{b}}, \apjl, 885, L24

\bibitem[{{CHIME/FRB Collaboration} {et~al.}(2020{\natexlab{a}}){CHIME/FRB
  Collaboration}, {Amiri}, {Andersen}, {Bandura}, {Bhardwaj}, {Boyle}, {Brar},
  {Chawla}, {Chen}, {Cliche}, {Cubranic}, {Deng}, {Denman}, {Dobbs}, {Dong},
  {Fandino}, {Fonseca}, {Gaensler}, {Giri}, {Good}, {Halpern}, {Hessels},
  {Hill}, {H{\"o}fer}, {Josephy}, {Kania}, {Karuppusamy}, {Kaspi}, {Keimpema},
  {Kirsten}, {Landecker}, {Lang}, {Leung}, {Li}, {Lin}, {Marcote}, {Masui},
  {McKinven}, {Mena-Parra}, {Merryfield}, {Michilli}, {Milutinovic},
  {Mirhosseini}, {Naidu}, {Newburgh}, {Ng}, {Nimmo}, {Paragi}, {Patel}, {Pen},
  {Pinsonneault-Marotte}, {Pleunis}, {Rafiei-Ravandi}, {Rahman}, {Ransom},
  {Renard}, {Sanghavi}, {Scholz}, {Shaw}, {Shin}, {Siegel}, {Singh}, {Smegal},
  {Smith}, {Stairs}, {Tendulkar}, {Tretyakov}, {Vanderlinde}, {Wang}, {Wang},
  {Wulf}, {Yadav}, \& {Zwaniga}}]{Amiri:2020gno}
{CHIME/FRB Collaboration}, {Amiri}, M., {Andersen}, B.~C., {et~al.}
  2020{\natexlab{a}}, \nat, 582, 351

\bibitem[{{CHIME/FRB Collaboration} {et~al.}(2020{\natexlab{b}}){CHIME/FRB
  Collaboration}, {:}, {Andersen}, {Band ura}, {Bhardwaj}, {Bij}, {Boyce},
  {Boyle}, {Brar}, {Cassanelli}, {Chawla}, {Chen}, {Cliche}, {Cook},
  {Cubranic}, {Curtin}, {Denman}, {Dobbs}, {Dong}, {Fandino}, {Fonseca},
  {Gaensler}, {Giri}, {Good}, {Halpern}, {Hill}, {Hinshaw}, {H{\"o}fer},
  {Josephy}, {Kania}, {Kaspi}, {Landecker}, {Leung}, {Li}, {Lin}, {Masui},
  {Mckinven}, {Mena-Parra}, {Merryfield}, {Meyers}, {Michilli}, {Milutinovic},
  {Mirhosseini}, {M{\"u}nchmeyer}, {Naidu}, {Newburgh}, {Ng}, {Patel}, {Pen},
  {Pinsonneault-Marotte}, {Pleunis}, {Quine}, {Rafiei-Ravandi}, {Rahman},
  {Ransom}, {Renard}, {Sanghavi}, {Scholz}, {Shaw}, {Shin}, {Siegel}, {Singh},
  {Smegal}, {Smith}, {Stairs}, {Tan}, {Tendulkar}, {Tretyakov}, {Vanderlinde},
  {Wang}, {Wulf}, \& {Zwaniga}}]{2020SGR}
{CHIME/FRB Collaboration}, {:}, {Andersen}, B.~C., {et~al.} 2020{\natexlab{b}},
  arXiv e-prints, arXiv:2005.10324

\bibitem[{{Clark} {et~al.}(2018){Clark}, {Verstocken}, {Bianchi}, {Fritz},
  {Viaene}, {Smith}, {Baes}, {Casasola}, {Cassara}, {Davies}, {De Looze}, {De
  Vis}, {Evans}, {Galametz}, {Jones}, {Lianou}, {Madden}, {Mosenkov}, \&
  {Xilouris}}]{clarke2018A&A}
{Clark}, C.~J.~R., {Verstocken}, S., {Bianchi}, S., {et~al.} 2018, \aap, 609,
  A37

\bibitem[{Condon {et~al.}(1998)Condon, Cotton, Greisen, Yin, Perley, Taylor, \&
  Broderick}]{condon1998nrao}
Condon, J.~J., Cotton, W., Greisen, E., {et~al.} 1998, The Astronomical
  Journal, 115, 1693

\bibitem[{{Condon} {et~al.}(1998){Condon}, {Yin}, {Thuan}, \&
  {Boller}}]{condom1998AJ}
{Condon}, J.~J., {Yin}, Q.~F., {Thuan}, T.~X., \& {Boller}, T. 1998, \aj, 116,
  2682

\bibitem[{{Conroy} {et~al.}(2009){Conroy}, {Gunn}, \& {White}}]{Conroy2009}
{Conroy}, C., {Gunn}, J.~E., \& {White}, M. 2009, \apj, 699, 486

\bibitem[{Cordes \& Lazio(2002)}]{cordes2002ne2001}
Cordes, J.~M., \& Lazio, T. J.~W. 2002, arXiv preprint astro-ph/0207156

\bibitem[{{Cruces} {et~al.}(2020){Cruces}, {Spitler}, {Scholz}, {Lynch},
  {Seymour}, {Hessels}, {Gouiff{\'e}s}, {Hilmarsson}, {Kramer}, \&
  {Munjal}}]{2021MNRAS.500..448C}
{Cruces}, M., {Spitler}, L.~G., {Scholz}, P., {et~al.} 2020, \mnras, 500, 448

\bibitem[{{Dale} \& {Helou}(2002)}]{dale2002}
{Dale}, D.~A., \& {Helou}, G. 2002, \apj, 576, 159

\bibitem[{{Davies} {et~al.}(2017){Davies}, {Huynh}, {Hopkins}, {Seymour},
  {Driver}, {Robotham}, {Baldry}, {Bland-Hawthorn}, {Bourne}, {Bremer},
  {Brown}, {Brough}, {Cluver}, {Grootes}, {Jarvis}, {Loveday}, {Moffet},
  {Owers}, {Phillipps}, {Sadler}, {Wang}, {Wilkins}, \&
  {Wright}}]{davis2017MNRAS}
{Davies}, L.~J.~M., {Huynh}, M.~T., {Hopkins}, A.~M., {et~al.} 2017, \mnras,
  466, 2312

\bibitem[{{de Vaucouleurs} {et~al.}(1991){de Vaucouleurs}, {de Vaucouleurs},
  {Corwin}, {Buta}, {Paturel}, \& {Fouque}}]{de1991third}
{de Vaucouleurs}, G., {de Vaucouleurs}, A., {Corwin}, Herold~G., J., {et~al.}
  1991, {Third Reference Catalogue of Bright Galaxies} (Springer)

\bibitem[{{Dey} {et~al.}(2019){Dey}, {Schlegel}, {Lang}, {Blum}, {Burleigh},
  {Fan}, {Findlay}, {Finkbeiner}, {Herrera}, {Juneau}, {Landriau}, {Levi},
  {McGreer}, {Meisner}, {Myers}, {Moustakas}, {Nugent}, {Patej}, {Schlafly},
  {Walker}, {Valdes}, {Weaver}, {Y{\`e}che}, {Zou}, {Zhou}, {Abareshi},
  {Abbott}, {Abolfathi}, {Aguilera}, {Alam}, {Allen}, {Alvarez}, {Annis},
  {Ansarinejad}, {Aubert}, {Beechert}, {Bell}, {BenZvi}, {Beutler}, {Bielby},
  {Bolton}, {Brice{\~n}o}, {Buckley-Geer}, {Butler}, {Calamida}, {Carlberg},
  {Carter}, {Casas}, {Castander}, {Choi}, {Comparat}, {Cukanovaite}, {Delubac},
  {DeVries}, {Dey}, {Dhungana}, {Dickinson}, {Ding}, {Donaldson}, {Duan},
  {Duckworth}, {Eftekharzadeh}, {Eisenstein}, {Etourneau}, {Fagrelius},
  {Farihi}, {Fitzpatrick}, {Font-Ribera}, {Fulmer}, {G{\"a}nsicke},
  {Gaztanaga}, {George}, {Gerdes}, {Gontcho}, {Gorgoni}, {Green}, {Guy},
  {Harmer}, {Hernand ez}, {Honscheid}, {Huang}, {James}, {Jannuzi}, {Jiang},
  {Joyce}, {Karcher}, {Karkar}, {Kehoe}, {Kneib}, {Kueter-Young}, {Lan},
  {Lauer}, {Le Guillou}, {Le Van Suu}, {Lee}, {Lesser}, {Perreault Levasseur},
  {Li}, {Mann}, {Marshall}, {Mart{\'\i}nez-V{\'a}zquez}, {Martini}, {du Mas des
  Bourboux}, {McManus}, {Meier}, {M{\'e}nard}, {Metcalfe},
  {Mu{\~n}oz-Guti{\'e}rrez}, {Najita}, {Napier}, {Narayan}, {Newman}, {Nie},
  {Nord}, {Norman}, {Olsen}, {Paat}, {Palanque-Delabrouille}, {Peng},
  {Poppett}, {Poremba}, {Prakash}, {Rabinowitz}, {Raichoor}, {Rezaie},
  {Robertson}, {Roe}, {Ross}, {Ross}, {Rudnick}, {Safonova}, {Saha},
  {S{\'a}nchez}, {Savary}, {Schweiker}, {Scott}, {Seo}, {Shan}, {Silva},
  {Slepian}, {Soto}, {Sprayberry}, {Staten}, {Stillman}, {Stupak}, {Summers},
  {Sien Tie}, {Tirado}, {Vargas-Maga{\~n}a}, {Vivas}, {Wechsler}, {Williams},
  {Yang}, {Yang}, {Yapici}, {Zaritsky}, {Zenteno}, {Zhang}, {Zhang}, {Zhou}, \&
  {Zhou}}]{DESI-2019}
{Dey}, A., {Schlegel}, D.~J., {Lang}, D., {et~al.} 2019, \aj, 157, 168

\bibitem[{Dolag {et~al.}(2015)Dolag, Gaensler, Beck, \&
  Beck}]{dolag2015constraints}
Dolag, K., Gaensler, B.~M., Beck, A.~M., \& Beck, M.~C. 2015, Monthly Notices
  of the Royal Astronomical Society, 451, 4277

\bibitem[{{Draine} \& {Li}(2007)}]{draine2007}
{Draine}, B.~T., \& {Li}, A. 2007, \apj, 657, 810

\bibitem[{Driver {et~al.}(2016)Driver, Andrews, Davies, Robotham, Wright,
  Windhorst, Cohen, Emig, Jansen, \& Dunne}]{driver2016measurements}
Driver, S.~P., Andrews, S.~K., Davies, L.~J., {et~al.} 2016, The Astrophysical
  Journal, 827, 108

\bibitem[{{Fong} {et~al.}(2021){Fong}, {Dong}, {Leja}, {Bhandari}, {Day},
  {Deller}, {Kumar}, {Prochaska}, {Scott}, {Bannister}, {Eftekhari}, {Gordon},
  {Heintz}, {James}, {Kilpatrick}, {Mahony}, {Rouco Escorial}, {Ryder},
  {Shannon}, \& {Tejos}}]{2021arXiv210611993F}
{Fong}, W.-f., {Dong}, Y., {Leja}, J., {et~al.} 2021, arXiv e-prints,
  arXiv:2106.11993

\bibitem[{{Fonseca} {et~al.}(2020){Fonseca}, {Andersen}, {Bhardwaj}, {Chawla},
  {Good}, {Josephy}, {Kaspi}, {Masui}, {Mckinven}, {Michilli}, {Pleunis},
  {Shin}, {Tendulkar}, {Bandura}, {Boyle}, {Brar}, {Cassanelli}, {Cubranic},
  {Dobbs}, {Dong}, {Gaensler}, {Hinshaw}, {Land ecker}, {Leung}, {Li}, {Lin},
  {Mena-Parra}, {Merryfield}, {Naidu}, {Ng}, {Patel}, {Pen}, {Rafiei-Ravandi},
  {Rahman}, {Ransom}, {Scholz}, {Smith}, {Stairs}, {Vanderlinde}, {Yadav}, \&
  {Zwaniga}}]{fab+20}
{Fonseca}, E., {Andersen}, B.~C., {Bhardwaj}, M., {et~al.} 2020, \apjl,
  arXiv:2001.03595

\bibitem[{{Foreman-Mackey} {et~al.}(2013){Foreman-Mackey}, {Hogg}, {Lang}, \&
  {Goodman}}]{fm2013}
{Foreman-Mackey}, D., {Hogg}, D.~W., {Lang}, D., \& {Goodman}, J. 2013, \pasp,
  125, 306

\bibitem[{{Fruchter} {et~al.}(2006){Fruchter}, {Levan}, {Strolger},
  {Vreeswijk}, {Thorsett}, {Bersier}, {Burud}, {Castro Cer{\'o}n},
  {Castro-Tirado}, {Conselice}, {Dahlen}, {Ferguson}, {Fynbo}, {Garnavich},
  {Gibbons}, {Gorosabel}, {Gull}, {Hjorth}, {Holland}, {Kouveliotou}, {Levay},
  {Livio}, {Metzger}, {Nugent}, {Petro}, {Pian}, {Rhoads}, {Riess}, {Sahu},
  {Smette}, {Tanvir}, {Wijers}, \& {Woosley}}]{fruchter2006}
{Fruchter}, A.~S., {Levan}, A.~J., {Strolger}, L., {et~al.} 2006, \nat, 441,
  463

\bibitem[{Fukugita {et~al.}(1998)Fukugita, Hogan, \&
  Peebles}]{fukugita1998cosmic}
Fukugita, M., Hogan, C., \& Peebles, P. 1998, The Astrophysical Journal, 503,
  518

\bibitem[{{Fullmer} \& {Londsale}(1995)}]{Fullmer1995}
{Fullmer}, L., \& {Londsale}, C.~J. 1995, VizieR Online Data Catalog, VII/113

\bibitem[{Gelman {et~al.}(2013)Gelman, Carlin, Stern, Dunson, Vehtari, \&
  Rubin}]{gelman2013}
Gelman, A., Carlin, J.~B., Stern, H.~S., {et~al.} 2013, Bayesian data analysis
  (CRC press)

\bibitem[{Goldstein {et~al.}(2021)Goldstein, Cleveland, \&
  Kocevski}]{GbmDataTools}
Goldstein, A., Cleveland, W.~H., \& Kocevski, D. 2021, Fermi GBM Data Tools:
  v1.1.0, , .
\newblock \url{https://fermi.gsfc.nasa.gov/ssc/data/analysis/gbm}

\bibitem[{{G{\'o}mez-Gonz{\'a}lez} {et~al.}(2016){G{\'o}mez-Gonz{\'a}lez},
  {Mayya}, \& {Rosa-Gonz{\'a}lez}}]{gtcmos}
{G{\'o}mez-Gonz{\'a}lez}, V.~M.~A., {Mayya}, Y.~D., \& {Rosa-Gonz{\'a}lez}, D.
  2016, \mnras, 460, 1555

\bibitem[{{G{\'o}mez-Velarde} {et~al.}(2016){G{\'o}mez-Velarde},
  {Garc{\'\i}a-Alvarez}, \& {Cabrerra-Lavers}}]{mask2}
{G{\'o}mez-Velarde}, G., {Garc{\'\i}a-Alvarez}, D., \& {Cabrerra-Lavers}, A.
  2016, in Astronomical Society of the Pacific Conference Series, Vol. 507,
  Multi-Object Spectroscopy in the Next Decade: Big Questions, Large Surveys,
  and Wide Fields, ed. I.~{Skillen}, M.~{Balcells}, \& S.~{Trager}, 191

\bibitem[{{Gonz{\'a}lez-Serrano} {et~al.}(2004){Gonz{\'a}lez-Serrano},
  {S{\'a}nchez-Portal}, {Casta{\~n}eda}, {Quirk}, {de Miguel}, {Aguiar}, \&
  {Cepa}}]{mask1}
{Gonz{\'a}lez-Serrano}, J.~I., {S{\'a}nchez-Portal}, M., {Casta{\~n}eda}, H.,
  {et~al.} 2004, Experimental Astronomy, 18, 65

\bibitem[{{Goodman} \& {Weare}(2010)}]{gw2010}
{Goodman}, J., \& {Weare}, J. 2010, Communications in Applied Mathematics and
  Computational Science, 5, 65

\bibitem[{{Gourdji} {et~al.}(2019){Gourdji}, {Michilli}, {Spitler}, {Hessels},
  {Seymour}, {Cordes}, \& {Chatterjee}}]{gourdji2019ApJ}
{Gourdji}, K., {Michilli}, D., {Spitler}, L.~G., {et~al.} 2019, \apjl, 877, L19

\bibitem[{{Greco} {et~al.}(2018){Greco}, {Greene}, {Strauss}, {Macarthur},
  {Flowers}, {Goulding}, {Huang}, {Kim}, {Komiyama}, {Leauthaud}, {Leisman},
  {Lupton}, {Sif{\'o}n}, \& {Wang}}]{Greco2018ApJ}
{Greco}, J.~P., {Greene}, J.~E., {Strauss}, M.~A., {et~al.} 2018, \apj, 857,
  104

\bibitem[{{Haakonsen} \& {Rutledge}(2009)}]{Haakonsen2009ApJS}
{Haakonsen}, C.~B., \& {Rutledge}, R.~E. 2009, \apjs, 184, 138

\bibitem[{Hafen {et~al.}(2019)Hafen, Faucher-Gigu{\`e}re,
  Angl{\'e}s-Alc{\'a}zar, Stern, Kere{\v{s}}, Hummels, Esmerian,
  Garrison-Kimmel, El-Badry, Wetzel, {et~al.}}]{hafen2019origins}
Hafen, Z., Faucher-Gigu{\`e}re, C.-A., Angl{\'e}s-Alc{\'a}zar, D., {et~al.}
  2019, Monthly Notices of the Royal Astronomical Society, 488, 1248

\bibitem[{Hancock {et~al.}(2012)Hancock, Murphy, Gaensler, Hopkins, \&
  Curran}]{hancock2012compact}
Hancock, P.~J., Murphy, T., Gaensler, B.~M., Hopkins, A., \& Curran, J.~R.
  2012, Monthly Notices of the Royal Astronomical Society, 422, 1812

\bibitem[{Hancock {et~al.}(2018)Hancock, Trott, \&
  Hurley-Walker}]{hancock2018source}
Hancock, P.~J., Trott, C.~M., \& Hurley-Walker, N. 2018, Publications of the
  Astronomical Society of Australia, 35

\bibitem[{Harris {et~al.}(2020)Harris, Millman, van~der Walt, Gommers,
  Virtanen, Cournapeau, Wieser, Taylor, Berg, Smith, Kern, Picus, Hoyer, van
  Kerkwijk, Brett, Haldane, del R{\'{i}}o, Wiebe, Peterson,
  G{\'{e}}rard-Marchant, Sheppard, Reddy, Weckesser, Abbasi, Gohlke, \&
  Oliphant}]{harris2020array}
Harris, C.~R., Millman, K.~J., van~der Walt, S.~J., {et~al.} 2020, Nature, 585,
  357.
\newblock \url{https://doi.org/10.1038/s41586-020-2649-2}

\bibitem[{Heintz {et~al.}(2020)Heintz, Prochaska, Simha, Platts, Fong, Tejos,
  Ryder, Aggerwal, Bhandari, Day, {et~al.}}]{heintz2020host}
Heintz, K.~E., Prochaska, J.~X., Simha, S., {et~al.} 2020, arXiv preprint
  arXiv:2009.10747

\bibitem[{{Hinton} {et~al.}(2016){Hinton}, {Davis}, {Lidman}, {Glazebrook}, \&
  {Lewis}}]{Hinton}
{Hinton}, S.~R., {Davis}, T.~M., {Lidman}, C., {Glazebrook}, K., \& {Lewis},
  G.~F. 2016, Astronomy and Computing, 15, 61

\bibitem[{{Huchra} {et~al.}(1983){Huchra}, {Davis}, {Latham}, \&
  {Tonry}}]{huchra1983ApJS}
{Huchra}, J., {Davis}, M., {Latham}, D., \& {Tonry}, J. 1983, \apjs, 52, 89

\bibitem[{Hunter(2007)}]{Hunter:2007}
Hunter, J.~D. 2007, Computing in Science \& Engineering, 9, 90

\bibitem[{{Iglesias-P{\'a}ramo} {et~al.}(2004){Iglesias-P{\'a}ramo}, {Buat},
  {Donas}, {Boselli}, \& {Milliard}}]{Iglesias2004A&A}
{Iglesias-P{\'a}ramo}, J., {Buat}, V., {Donas}, J., {Boselli}, A., \&
  {Milliard}, B. 2004, \aap, 419, 109

\bibitem[{{Iglesias-P{\'a}ramo} {et~al.}(2006){Iglesias-P{\'a}ramo}, {Buat},
  {Takeuchi}, {Xu}, {Boissier}, {Boselli}, {Burgarella}, {Madore}, {Gil de
  Paz}, {Bianchi}, {Barlow}, {Byun}, {Donas}, {Forster}, {Friedman}, {Heckman},
  {Jelinski}, {Lee}, {Malina}, {Martin}, {Milliard}, {Morrissey}, {Neff},
  {Rich}, {Schiminovich}, {Seibert}, {Siegmund}, {Small}, {Szalay}, {Welsh}, \&
  {Wyder}}]{Iglesias2006}
{Iglesias-P{\'a}ramo}, J., {Buat}, V., {Takeuchi}, T.~T., {et~al.} 2006, \apjs,
  164, 38

\bibitem[{Intema {et~al.}(2017)Intema, Jagannathan, Mooley, \&
  Frail}]{intema2017gmrt}
Intema, H., Jagannathan, P., Mooley, K., \& Frail, D. 2017, Astronomy \&
  Astrophysics, 598, A78

\bibitem[{{Johnson} {et~al.}(2019){Johnson}, {Leja}, {Conroy}, \&
  {Speagle}}]{prospect2019}
{Johnson}, B.~D., {Leja}, J.~L., {Conroy}, C., \& {Speagle}, J.~S. 2019,
  {Prospector: Stellar population inference from spectra and SEDs},
  Astrophysics Source Code Library, ascl:1905.025

\bibitem[{{Joye} \& {Mandel}(2003)}]{2003ASPC..295..489J}
{Joye}, W.~A., \& {Mandel}, E. 2003, in Astronomical Society of the Pacific
  Conference Series, Vol. 295, Astronomical Data Analysis Software and Systems
  XII, ed. H.~E. {Payne}, R.~I. {Jedrzejewski}, \& R.~N. {Hook}, 489

\bibitem[{Kaaret {et~al.}(2020)Kaaret, Koutroumpa, Kuntz, Jahoda, Bluem,
  Gulick, Hodges-Kluck, LaRocca, Ringuette, \& Zajczyk}]{Kaaret2020}
Kaaret, P., Koutroumpa, D., Kuntz, K.~D., {et~al.} 2020, Nature Astronomy,
  doi:10.1038/s41550-020-01215-w.
\newblock \url{https://doi.org/10.1038/s41550-020-01215-w}

\bibitem[{{Kasliwal}(2011)}]{ptf10hjz}
{Kasliwal}, M. 2011, the Gravitational-wave Physics and Astronomy
  Workshop,January 26-29, 2011.
\newblock
  \url{http://www.gravity.phys.uwm.edu/conferences/gwpaw/program/talks.html}

\bibitem[{{Keane} {et~al.}(2016){Keane}, {Johnston}, {Bhandari}, {Barr},
  {Bhat}, {Burgay}, {Caleb}, {Flynn}, {Jameson}, {Kramer}, {Petroff},
  {Possenti}, {van Straten}, {Bailes}, {Burke-Spolaor}, {Eatough}, {Stappers},
  {Totani}, {Honma}, {Furusawa}, {Hattori}, {Morokuma}, {Niino}, {Sugai},
  {Terai}, {Tominaga}, {Yamasaki}, {Yasuda}, {Allen}, {Cooke}, {Jencson},
  {Kasliwal}, {Kaplan}, {Tingay}, {Williams}, {Wayth}, {Chandra}, {Perrodin},
  {Berezina}, {Mickaliger}, \& {Bassa}}]{Keane2016}
{Keane}, E.~F., {Johnston}, S., {Bhandari}, S., {et~al.} 2016, \nat, 530, 453

\bibitem[{{Keating} \& {Pen}(2020)}]{keating2020exploring}
{Keating}, L.~C., \& {Pen}, U.-L. 2020, \mnras, 496, L106

\bibitem[{{Kennicutt} {et~al.}(1994){Kennicutt}, {Tamblyn}, \&
  {Congdon}}]{kennicutt1994ApJ}
{Kennicutt}, Robert~C., J., {Tamblyn}, P., \& {Congdon}, C.~E. 1994, \apj, 435,
  22

\bibitem[{{Kirsten} {et~al.}(2021){Kirsten}, {Marcote}, {Nimmo}, {Hessels},
  {Bhardwaj}, {Tendulkar}, {Keimpema}, {Yang}, {Snelders}, {Scholz},
  {Pearlman}, {Law}, {Peters}, {Giroletti}, {Hewitt}, {Bach}, {Bezukovs},
  {Burgay}, {Buttaccio}, {Conway}, {Corongiu}, {Feiler}, {Forss{\'e}n},
  {Gawro{\'n}ski}, {Karuppusamy}, {Kharinov}, {Lindqvist}, {Maccaferri},
  {Melnikov}, {Ould-Boukattine}, {Paragi}, {Possenti}, {Surcis}, {Wang},
  {Yuan}, {Aggarwal}, {Anna-Thomas}, {Bower}, {Blaauw}, {Burke-Spolaor},
  {Cassanelli}, {Clarke}, {Fonseca}, {Gaensler}, {Gopinath}, {Kaspi}, {Kassim},
  {Lazio}, {Leung}, {Li}, {Lin}, {Masui}, {Mckinven}, {Michilli}, {Mikhailov},
  {Ng}, {Orbidans}, {Pen}, {Petroff}, {Rahman}, {Ransom}, {Shin}, {Smith},
  {Stairs}, \& {Vlemmings}}]{Kristan2021arXiv}
{Kirsten}, F., {Marcote}, B., {Nimmo}, K., {et~al.} 2021, arXiv e-prints,
  arXiv:2105.11445

\bibitem[{{Klein} {et~al.}(2018){Klein}, {Lisenfeld}, \&
  {Verley}}]{2018A&A...611A..55K}
{Klein}, U., {Lisenfeld}, U., \& {Verley}, S. 2018, \aap, 611, A55

\bibitem[{{Klypin} {et~al.}(2016){Klypin}, {Yepes}, {Gottl{\"o}ber}, {Prada},
  \& {He{\ss}}}]{klypin2016}
{Klypin}, A., {Yepes}, G., {Gottl{\"o}ber}, S., {Prada}, F., \& {He{\ss}}, S.
  2016, \mnras, 457, 4340

\bibitem[{{Kourkchi} \& {Tully}(2017)}]{Kourkchi2017}
{Kourkchi}, E., \& {Tully}, R.~B. 2017, \apj, 843, 16

\bibitem[{{Kwiatkowski}(2015)}]{Kwiatkowski2015}
{Kwiatkowski}, D. 2015, arXiv e-prints, arXiv:1512.00678

\bibitem[{Lacy {et~al.}(2016)Lacy, Baum, Chandler, Chatterjee, Murphy, Myers,
  {et~al.}}]{lacy2016vla}
Lacy, M., Baum, S.~A., Chandler, C.~J., {et~al.} 2016, AAS, 227, 324

\bibitem[{{Leitherer}(2005)}]{leitherer2005}
{Leitherer}, C. 2005, in American Institute of Physics Conference Series, Vol.
  761, The Spectral Energy Distributions of Gas-Rich Galaxies: Confronting
  Models with Data, ed. C.~C. {Popescu} \& R.~J. {Tuffs}, 39--58

\bibitem[{{Leja} {et~al.}(2017){Leja}, {Johnson}, {Conroy}, {van Dokkum}, \&
  {Byler}}]{Leja2017}
{Leja}, J., {Johnson}, B.~D., {Conroy}, C., {van Dokkum}, P.~G., \& {Byler}, N.
  2017, \apj, 837, 170

\bibitem[{Li {et~al.}(2020)Li, Lin, Xiong, Ge, Li, Li, Lu, Zhang, Tuo, Nang,
  {et~al.}}]{li2020identification}
Li, C., Lin, L., Xiong, S., {et~al.} 2020, arXiv preprint arXiv:2005.11071

\bibitem[{{Li} \& {Zhang}(2020)}]{li2020}
{Li}, Y., \& {Zhang}, B. 2020, \apjl, 899, L6

\bibitem[{{Li} {et~al.}(2019){Li}, {Zhang}, {Nagamine}, \& {Shi}}]{Li2019ApJ}
{Li}, Y., {Zhang}, B., {Nagamine}, K., \& {Shi}, J. 2019, \apjl, 884, L26

\bibitem[{{Lim} {et~al.}(2017){Lim}, {Mo}, {Lu}, {Wang}, \& {Yang}}]{lim2017}
{Lim}, S.~H., {Mo}, H.~J., {Lu}, Y., {Wang}, H., \& {Yang}, X. 2017, \mnras,
  470, 2982

\bibitem[{{Lorimer} {et~al.}(2007){Lorimer}, {Bailes}, {McLaughlin},
  {Narkevic}, \& {Crawford}}]{lbm+07}
{Lorimer}, D.~R., {Bailes}, M., {McLaughlin}, M.~A., {Narkevic}, D.~J., \&
  {Crawford}, F. 2007, Science, 318, 777

\bibitem[{{Lu} \& {Kumar}(2016)}]{lu2016MNRAS}
{Lu}, W., \& {Kumar}, P. 2016, \mnras, 461, L122

\bibitem[{{Lu} \& {Piro}(2019)}]{lu2019}
{Lu}, W., \& {Piro}, A.~L. 2019, \apj, 883, 40

\bibitem[{{Luo} {et~al.}(2020){Luo}, {Men}, {Lee}, {Wang}, {Lorimer}, \&
  {Zhang}}]{luo2020}
{Luo}, R., {Men}, Y., {Lee}, K., {et~al.} 2020, \mnras, 494, 665

\bibitem[{{Lyubarsky}(2021)}]{2021yuri}
{Lyubarsky}, Y. 2021, Universe, 7, 56

\bibitem[{Macquart {et~al.}(2020)Macquart, Prochaska, McQuinn, Bannister,
  Bhandari, Day, Deller, Ekers, James, Marnoch, {et~al.}}]{macquart2020census}
Macquart, J.-P., Prochaska, J., McQuinn, M., {et~al.} 2020, Nature, 581, 391

\bibitem[{Maller \& Bullock(2004)}]{maller2004multiphase}
Maller, A.~H., \& Bullock, J.~S. 2004, Monthly Notices of the Royal
  Astronomical Society, 355, 694

\bibitem[{{Mannings} {et~al.}(2020){Mannings}, {Fong}, {Simha}, {Prochaska},
  {Rafelski}, {Kilpatrick}, {Tejos}, {Heintz}, {Bhandari}, {Day}, {Deller},
  {Ryder}, {Shannon}, \& {Tendulkar}}]{Mannings+21}
{Mannings}, A.~G., {Fong}, W.-f., {Simha}, S., {et~al.} 2020, arXiv e-prints,
  arXiv:2012.11617

\bibitem[{{Maoz}(2007)}]{2007MNRAS.377.1696M}
{Maoz}, D. 2007, \mnras, 377, 1696

\bibitem[{{Marvil} {et~al.}(2015){Marvil}, {Owen}, \& {Eilek}}]{Marvil2015AJ}
{Marvil}, J., {Owen}, F., \& {Eilek}, J. 2015, \aj, 149, 32

\bibitem[{{Masters} {et~al.}(2014){Masters}, {Crook}, {Hong}, {Jarrett},
  {Koribalski}, {Macri}, {Springob}, \& {Staveley-Smith}}]{2014MNRAS.443.1044M}
{Masters}, K.~L., {Crook}, A., {Hong}, T., {et~al.} 2014, \mnras, 443, 1044

\bibitem[{{McCall} {et~al.}(1985){McCall}, {Rybski}, \&
  {Shields}}]{mccall1985ApJS}
{McCall}, M.~L., {Rybski}, P.~M., \& {Shields}, G.~A. 1985, \apjs, 57, 1

\bibitem[{{McMullin} {et~al.}(2007){McMullin}, {Waters}, {Schiebel}, {Young},
  \& {Golap}}]{casa2007}
{McMullin}, J.~P., {Waters}, B., {Schiebel}, D., {Young}, W., \& {Golap}, K.
  2007, in Astronomical Society of the Pacific Conference Series, Vol. 376,
  Astronomical Data Analysis Software and Systems XVI, ed. R.~A. {Shaw},
  F.~{Hill}, \& D.~J. {Bell}, 127

\bibitem[{Mereghetti {et~al.}(2020)Mereghetti, Savchenko, Ferrigno, G{\"o}tz,
  Rigoselli, Tiengo, Bazzano, Bozzo, Coleiro, Courvoisier,
  {et~al.}}]{mereghetti2020integral}
Mereghetti, S., Savchenko, V., Ferrigno, C., {et~al.} 2020, arXiv preprint
  arXiv:2005.06335

\bibitem[{{Metzger} {et~al.}(2017){Metzger}, {Berger}, \&
  {Margalit}}]{metzger2017ApJ}
{Metzger}, B.~D., {Berger}, E., \& {Margalit}, B. 2017, \apj, 841, 14

\bibitem[{Metzger {et~al.}(2019)Metzger, Margalit, \& Sironi}]{metzger2019fast}
Metzger, B.~D., Margalit, B., \& Sironi, L. 2019, Monthly Notices of the Royal
  Astronomical Society, 485, 4091

\bibitem[{{Michilli} {et~al.}(2020){Michilli}, {Masui}, {Mckinven}, {Cubranic},
  {Bruneault}, {Brar}, {Patel}, {Boyle}, {Stairs}, {Renard}, {Bandura},
  {Berger}, {Breitman}, {Cassanelli}, {Dobbs}, {Kaspi}, {Leung}, {Mena-Parra},
  {Pleunis}, {Russell}, {Scholz}, {Siegel}, {Tendulkar}, \& {Vand
  erlinde}}]{Daniele2020}
{Michilli}, D., {Masui}, K.~W., {Mckinven}, R., {et~al.} 2020, arXiv e-prints,
  arXiv:2010.06748

\bibitem[{{Moster} {et~al.}(2013){Moster}, {Naab}, \& {White}}]{moster2013}
{Moster}, B.~P., {Naab}, T., \& {White}, S. D.~M. 2013, \mnras, 428, 3121

\bibitem[{{Nicastro} {et~al.}(2021){Nicastro}, {Guidorzi}, {Palazzi},
  {Zampieri}, {Turatto}, \& {Gardini}}]{nicastro2021}
{Nicastro}, L., {Guidorzi}, C., {Palazzi}, E., {et~al.} 2021, Universe, 7, 76

\bibitem[{{Nicholl} {et~al.}(2017{\natexlab{a}}){Nicholl}, {Williams},
  {Berger}, {Villar}, {Alexander}, {Eftekhari}, \& {Metzger}}]{nicholl2017}
{Nicholl}, M., {Williams}, P.~K.~G., {Berger}, E., {et~al.} 2017{\natexlab{a}},
  \apj, 843, 84

\bibitem[{{Nicholl} {et~al.}(2017{\natexlab{b}}){Nicholl}, {Williams},
  {Berger}, {Villar}, {Alexander}, {Eftekhari}, \& {Metzger}}]{nicholl2017ApJ}
---. 2017{\natexlab{b}}, \apj, 843, 84

\bibitem[{{Ocker} {et~al.}(2020){Ocker}, {Cordes}, \&
  {Chatterjee}}]{ocker2020ApJ}
{Ocker}, S.~K., {Cordes}, J.~M., \& {Chatterjee}, S. 2020, \apj, 897, 124

\bibitem[{{Oke}(1974)}]{std2}
{Oke}, J.~B. 1974, \apjs, 27, 21

\bibitem[{{Oke}(1990)}]{standard}
---. 1990, \aj, 99, 1621

\bibitem[{{Osterbrock} \& {Ferland}(2006)}]{Osterbrock2006}
{Osterbrock}, D.~E., \& {Ferland}, G.~J. 2006, {Astrophysics of gaseous nebulae
  and active galactic nuclei} (University Science Books)

\bibitem[{Palmer {et~al.}(2005)Palmer, Barthelmy, Gehrels, Kippen, Cayton,
  Kouveliotou, Eichler, Wijers, Woods, Granot, {et~al.}}]{palmer2005giant}
Palmer, D.~M., Barthelmy, S., Gehrels, N., {et~al.} 2005, Nature, 434, 1107

\bibitem[{{Pilyugin} \& {Grebel}(2016)}]{Pilyugin2016}
{Pilyugin}, L.~S., \& {Grebel}, E.~K. 2016, \mnras, 457, 3678

\bibitem[{{Planck Collaboration} {et~al.}(2016){Planck Collaboration}, {Ade},
  {Aghanim}, {Arnaud}, {Ashdown}, {Aumont}, {Baccigalupi}, {Banday},
  {Barreiro}, {Bartlett}, {Bartolo}, {Battaner}, {Battye}, {Benabed},
  {Beno{\^\i}t}, {Benoit-L{\'e}vy}, {Bernard}, {Bersanelli}, {Bielewicz},
  {Bock}, {Bonaldi}, {Bonavera}, {Bond}, {Borrill}, {Bouchet}, {Boulanger},
  {Bucher}, {Burigana}, {Butler}, {Calabrese}, {Cardoso}, {Catalano},
  {Challinor}, {Chamballu}, {Chary}, {Chiang}, {Chluba}, {Christensen},
  {Church}, {Clements}, {Colombi}, {Colombo}, {Combet}, {Coulais}, {Crill},
  {Curto}, {Cuttaia}, {Danese}, {Davies}, {Davis}, {de Bernardis}, {de Rosa},
  {de Zotti}, {Delabrouille}, {D{\'e}sert}, {Di Valentino}, {Dickinson},
  {Diego}, {Dolag}, {Dole}, {Donzelli}, {Dor{\'e}}, {Douspis}, {Ducout},
  {Dunkley}, {Dupac}, {Efstathiou}, {Elsner}, {En{\ss}lin}, {Eriksen},
  {Farhang}, {Fergusson}, {Finelli}, {Forni}, {Frailis}, {Fraisse},
  {Franceschi}, {Frejsel}, {Galeotta}, {Galli}, {Ganga}, {Gauthier}, {Gerbino},
  {Ghosh}, {Giard}, {Giraud-H{\'e}raud}, {Giusarma}, {Gjerl{\o}w},
  {Gonz{\'a}lez-Nuevo}, {G{\'o}rski}, {Gratton}, {Gregorio}, {Gruppuso},
  {Gudmundsson}, {Hamann}, {Hansen}, {Hanson}, {Harrison}, {Helou},
  {Henrot-Versill{\'e}}, {Hern{\'a}ndez-Monteagudo}, {Herranz}, {Hildebrand t},
  {Hivon}, {Hobson}, {Holmes}, {Hornstrup}, {Hovest}, {Huang}, {Huffenberger},
  {Hurier}, {Jaffe}, {Jaffe}, {Jones}, {Juvela}, {Keih{\"a}nen}, {Keskitalo},
  {Kisner}, {Kneissl}, {Knoche}, {Knox}, {Kunz}, {Kurki-Suonio}, {Lagache},
  {L{\"a}hteenm{\"a}ki}, {Lamarre}, {Lasenby}, {Lattanzi}, {Lawrence}, {Leahy},
  {Leonardi}, {Lesgourgues}, {Levrier}, {Lewis}, {Liguori}, {Lilje},
  {Linden-V{\o}rnle}, {L{\'o}pez-Caniego}, {Lubin}, {Mac{\'\i}as-P{\'e}rez},
  {Maggio}, {Maino}, {Mandolesi}, {Mangilli}, {Marchini}, {Maris}, {Martin},
  {Martinelli}, {Mart{\'\i}nez-Gonz{\'a}lez}, {Masi}, {Matarrese}, {McGehee},
  {Meinhold}, {Melchiorri}, {Melin}, {Mendes}, {Mennella}, {Migliaccio},
  {Millea}, {Mitra}, {Miville-Desch{\^e}nes}, {Moneti}, {Montier}, {Morgante},
  {Mortlock}, {Moss}, {Munshi}, {Murphy}, {Naselsky}, {Nati}, {Natoli},
  {Netterfield}, {N{\o}rgaard-Nielsen}, {Noviello}, {Novikov}, {Novikov},
  {Oxborrow}, {Paci}, {Pagano}, {Pajot}, {Paladini}, {Paoletti}, {Partridge},
  {Pasian}, {Patanchon}, {Pearson}, {Perdereau}, {Perotto}, {Perrotta},
  {Pettorino}, {Piacentini}, {Piat}, {Pierpaoli}, {Pietrobon}, {Plaszczynski},
  {Pointecouteau}, {Polenta}, {Popa}, {Pratt}, {Pr{\'e}zeau}, {Prunet},
  {Puget}, {Rachen}, {Reach}, {Rebolo}, {Reinecke}, {Remazeilles}, {Renault},
  {Renzi}, {Ristorcelli}, {Rocha}, {Rosset}, {Rossetti}, {Roudier},
  {Rouill{\'e} d'Orfeuil}, {Rowan-Robinson}, {Rubi{\~n}o-Mart{\'\i}n},
  {Rusholme}, {Said}, {Salvatelli}, {Salvati}, {Sandri}, {Santos},
  {Savelainen}, {Savini}, {Scott}, {Seiffert}, {Serra}, {Shellard}, {Spencer},
  {Spinelli}, {Stolyarov}, {Stompor}, {Sudiwala}, {Sunyaev}, {Sutton},
  {Suur-Uski}, {Sygnet}, {Tauber}, {Terenzi}, {Toffolatti}, {Tomasi},
  {Tristram}, {Trombetti}, {Tucci}, {Tuovinen}, {T{\"u}rler}, {Umana},
  {Valenziano}, {Valiviita}, {Van Tent}, {Vielva}, {Villa}, {Wade}, {Wandelt},
  {Wehus}, {White}, {White}, {Wilkinson}, {Yvon}, {Zacchei}, \&
  {Zonca}}]{2016Aplanck}
{Planck Collaboration}, {Ade}, P.~A.~R., {Aghanim}, N., {et~al.} 2016, \aap,
  594, A13

\bibitem[{{Platts} {et~al.}(2018){Platts}, {Weltman}, {Walters}, {Tendulkar},
  {Gordin}, \& {Kandhai}}]{pww+18}
{Platts}, E., {Weltman}, A., {Walters}, A., {et~al.} 2018, arXiv e-prints,
  arXiv:1810.05836

\bibitem[{{Rajwade} {et~al.}(2020){Rajwade}, {Mickaliger}, {Stappers},
  {Morello}, {Agarwal}, {Bassa}, {Breton}, {Caleb}, {Karastergiou}, {Keane}, \&
  {Lorimer}}]{Rajwade:2020uat}
{Rajwade}, K.~M., {Mickaliger}, M.~B., {Stappers}, B.~W., {et~al.} 2020,
  \mnras, 495, 3551

\bibitem[{{Ravi} {et~al.}(2021){Ravi}, {Law}, {Li}, {Aggarwal},
  {Burke-Spolaor}, {Connor}, {Lazio}, {Simard}, {Somalwar}, \&
  {Tendulkar}}]{2021arXiv210609710R}
{Ravi}, V., {Law}, C.~J., {Li}, D., {et~al.} 2021, arXiv e-prints,
  arXiv:2106.09710

\bibitem[{Rengelink {et~al.}(1997)Rengelink, Tang, De~Bruyn, Miley, Bremer,
  Roettgering, \& Bremer}]{rengelink1997westerbork}
Rengelink, R., Tang, Y., De~Bruyn, A., {et~al.} 1997, Astronomy and
  Astrophysics Supplement Series, 124, 259

\bibitem[{{Resmi} {et~al.}(2020){Resmi}, {Vink}, \&
  {Ishwara-Chandra}}]{resmi2020}
{Resmi}, L., {Vink}, J., \& {Ishwara-Chandra}, C.~H. 2020, arXiv e-prints,
  arXiv:2010.14334

\bibitem[{Ridnaia {et~al.}(2020)Ridnaia, Svinkin, Frederiks, Bykov, Popov,
  Aptekar, Golenetskii, Lysenko, Tsvetkova, Ulanov,
  {et~al.}}]{ridnaia2020peculiar}
Ridnaia, A., Svinkin, D., Frederiks, D., {et~al.} 2020, arXiv preprint
  arXiv:2005.11178

\bibitem[{{Robitaille} \& {Bressert}(2012)}]{2012ascl.soft08017R}
{Robitaille}, T., \& {Bressert}, E. 2012, {APLpy: Astronomical Plotting Library
  in Python}, , , ascl:1208.017

\bibitem[{{Salo} {et~al.}(2015){Salo}, {Laurikainen}, {Laine}, {Comer{\'o}n},
  {Gadotti}, {Buta}, {Sheth}, {Zaritsky}, {Ho}, {Knapen}, {Athanassoula},
  {Bosma}, {Laine}, {Cisternas}, {Kim}, {Mu{\~n}oz-Mateos}, {Regan}, {Hinz},
  {Gil de Paz}, {Menendez-Delmestre}, {Mizusawa}, {Erroz-Ferrer}, {Meidt}, \&
  {Querejeta}}]{salo2015}
{Salo}, H., {Laurikainen}, E., {Laine}, J., {et~al.} 2015, \apjs, 219, 4

\bibitem[{{Schlafly} \& {Finkbeiner}(2011)}]{schlafly2011ApJ}
{Schlafly}, E.~F., \& {Finkbeiner}, D.~P. 2011, \apj, 737, 103

\bibitem[{{Schneider} {et~al.}(1992){Schneider}, {Thuan}, {Mangum}, \&
  {Miller}}]{1992ApJS...81....5S}
{Schneider}, S.~E., {Thuan}, T.~X., {Mangum}, J.~G., \& {Miller}, J. 1992,
  \apjs, 81, 5

\bibitem[{Scholz {et~al.}(2020)Scholz, Cook, Cruces, Hessels, Kaspi, Majid,
  Naidu, Pearlman, Spitler, Bandura, {et~al.}}]{scholz2020simultaneous}
Scholz, P., Cook, A., Cruces, M., {et~al.} 2020, arXiv preprint
  arXiv:2004.06082

\bibitem[{{Serenelli} {et~al.}(2009){Serenelli}, {Basu}, {Ferguson}, \&
  {Asplund}}]{2009ApJ...705L.123S}
{Serenelli}, A.~M., {Basu}, S., {Ferguson}, J.~W., \& {Asplund}, M. 2009,
  \apjl, 705, L123

\bibitem[{{Simha} {et~al.}(2014){Simha}, {Weinberg}, {Conroy}, {Dave},
  {Fardal}, {Katz}, \& {Oppenheimer}}]{simha2014}
{Simha}, V., {Weinberg}, D.~H., {Conroy}, C., {et~al.} 2014, arXiv e-prints,
  arXiv:1404.0402

\bibitem[{{Tanoglidis} {et~al.}(2021){Tanoglidis}, {Drlica-Wagner}, {Wei},
  {Li}, {S{\'a}nchez}, {Zhang}, {Peter}, {Feldmeier-Krause}, {Prat}, {Casey},
  {Palmese}, {S{\'a}nchez}, {DeRose}, {Conselice}, {Gagnon}, {Abbott},
  {Aguena}, {Allam}, {Avila}, {Bechtol}, {Bertin}, {Bhargava}, {Brooks},
  {Burke}, {Rosell}, {Kind}, {Carretero}, {Chang}, {Costanzi}, {da Costa}, {De
  Vicente}, {Desai}, {Diehl}, {Doel}, {Eifler}, {Everett}, {Evrard},
  {Flaugher}, {Frieman}, {Garc{\'\i}a-Bellido}, {Gerdes}, {Gruendl},
  {Gschwend}, {Gutierrez}, {Hartley}, {Hollowood}, {Huterer}, {James},
  {Krause}, {Kuehn}, {Kuropatkin}, {Maia}, {March}, {Marshall}, {Menanteau},
  {Miquel}, {Ogando}, {Paz-Chinch{\'o}n}, {Romer}, {Roodman}, {Sanchez},
  {Scarpine}, {Serrano}, {Sevilla-Noarbe}, {Smith}, {Suchyta}, {Tarle},
  {Thomas}, {Tucker}, {Walker}, \& {DES Collaboration}}]{tanoglidis2021ApJS}
{Tanoglidis}, D., {Drlica-Wagner}, A., {Wei}, K., {et~al.} 2021, \apjs, 252, 18

\bibitem[{{Tauris} {et~al.}(2013){Tauris}, {Sanyal}, {Yoon}, \&
  {Langer}}]{Tauris2013A&A}
{Tauris}, T.~M., {Sanyal}, D., {Yoon}, S.~C., \& {Langer}, N. 2013, \aap, 558,
  A39

\bibitem[{{Taylor} {et~al.}(2014){Taylor}, {Cinabro}, {Dilday}, {Galbany},
  {Gupta}, {Kessler}, {Marriner}, {Nichol}, {Richmond}, {Schneider}, \&
  {Sollerman}}]{taylor2014}
{Taylor}, M., {Cinabro}, D., {Dilday}, B., {et~al.} 2014, \apj, 792, 135

\bibitem[{{Tempel} {et~al.}(2016){Tempel}, {Kipper}, {Tamm}, {Gramann},
  {Einasto}, {Sepp}, \& {Tuvikene}}]{tempel2016}
{Tempel}, E., {Kipper}, R., {Tamm}, A., {et~al.} 2016, \aap, 588, A14

\bibitem[{Tendulkar {et~al.}(2017)Tendulkar, Bassa, Cordes, Bower, Law,
  Chatterjee, Adams, Bogdanov, Burke-Spolaor, Butler,
  {et~al.}}]{tendulkar2017host}
Tendulkar, S.~P., Bassa, C., Cordes, J.~M., {et~al.} 2017, The Astrophysical
  Journal Letters, 834, L7

\bibitem[{{The CHIME/FRB Collaboration} {et~al.}(2021){The CHIME/FRB
  Collaboration}, {:}, {Amiri}, {Andersen}, {Bandura}, {Berger}, {Bhardwaj},
  {Boyce}, {Boyle}, {Brar}, {Breitman}, {Cassanelli}, {Chawla}, {Chen},
  {Cliche}, {Cook}, {Cubranic}, {Curtin}, {Deng}, {Dobbs}, {Fengqiu}, {Dong},
  {Eadie}, {Fandino}, {Fonseca}, {Gaensler}, {Giri}, {Good}, {Halpern}, {Hill},
  {Hinshaw}, {Josephy}, {Kaczmarek}, {Kader}, {Kania}, {Kaspi}, {Landecker},
  {Lang}, {Leung}, {Li}, {Lin}, {Masui}, {Mckinven}, {Mena-Parra},
  {Merryfield}, {Meyers}, {Michilli}, {Milutinovic}, {Mirhosseini},
  {M{\"u}nchmeyer}, {Naidu}, {Newburgh}, {Ng}, {Patel}, {Pen}, {Petroff},
  {Pinsonneault-Marotte}, {Pleunis}, {Rafiei-Ravandi}, {Rahman}, {Ransom},
  {Renard}, {Sanghavi}, {Scholz}, {Shaw}, {Shin}, {Siegel}, {Sikora}, {Singh},
  {Smith}, {Stairs}, {Tan}, {Tendulkar}, {Vanderlinde}, {Wang}, {Wulf}, \&
  {Zwaniga}}]{firstchimefrbcatalog2021}
{The CHIME/FRB Collaboration}, {:}, {Amiri}, M., {et~al.} 2021, arXiv e-prints,
  arXiv:2106.04352

\bibitem[{{Thornton} {et~al.}(2013){Thornton}, {Stappers}, {Bailes},
  {Barsdell}, {Bates}, {Bhat}, {Burgay}, {Burke-Spolaor}, {Champion}, {Coster},
  {D'Amico}, {Jameson}, {Johnston}, {Keith}, {Kramer}, {Levin}, {Milia}, {Ng},
  {Possenti}, \& {van Straten}}]{tsb+13}
{Thornton}, D., {Stappers}, B., {Bailes}, M., {et~al.} 2013, Science, 341, 53

\bibitem[{Tody(1986)}]{tody1986iraf}
Tody, D. 1986, in Instrumentation in astronomy VI, Vol. 627, International
  Society for Optics and Photonics, 733--748

\bibitem[{{Tody}(1993)}]{tody1993iraf}
{Tody}, D. 1993, in Astronomical Society of the Pacific Conference Series,
  Vol.~52, Astronomical Data Analysis Software and Systems II, ed. R.~J.
  {Hanisch}, R.~J.~V. {Brissenden}, \& J.~{Barnes}, 173

\bibitem[{{Truemper}(1982)}]{truemper1982rosat}
{Truemper}, J. 1982, Advances in Space Research, 2, 241

\bibitem[{{Tully}(2015)}]{tully2015AJ}
{Tully}, R.~B. 2015, \aj, 149, 171

\bibitem[{{Tully} {et~al.}(2016){Tully}, {Courtois}, \& {Sorce}}]{Tully2016}
{Tully}, R.~B., {Courtois}, H.~M., \& {Sorce}, J.~G. 2016, \aj, 152, 50

\bibitem[{{Tumlinson} {et~al.}(2017){Tumlinson}, {Peeples}, \&
  {Werk}}]{tumlinson2017}
{Tumlinson}, J., {Peeples}, M.~S., \& {Werk}, J.~K. 2017, \araa, 55, 389

\bibitem[{{Ueda} {et~al.}(2001){Ueda}, {Ishisaki}, {Takahashi}, {Makishima}, \&
  {Ohashi}}]{Ueda2001ApJS}
{Ueda}, Y., {Ishisaki}, Y., {Takahashi}, T., {Makishima}, K., \& {Ohashi}, T.
  2001, \apjs, 133, 1

\bibitem[{{van den Bergh} {et~al.}(2005){van den Bergh}, {Li}, \&
  {Filippenko}}]{Bergh2005PASP}
{van den Bergh}, S., {Li}, W., \& {Filippenko}, A.~V. 2005, \pasp, 117, 773

\bibitem[{{von Kienlin} {et~al.}(2020){von Kienlin}, {Meegan}, {Paciesas},
  {Bhat}, {Bissaldi}, {Briggs}, {Burns}, {Cleveland}, {Gibby}, {Giles},
  {Goldstein}, {Hamburg}, {Hui}, {Kocevski}, {Mailyan}, {Malacaria},
  {Poolakkil}, {Preece}, {Roberts}, {Veres}, \&
  {Wilson-Hodge}}]{2020ApJ...893...46V}
{von Kienlin}, A., {Meegan}, C.~A., {Paciesas}, W.~S., {et~al.} 2020, \apj,
  893, 46

\bibitem[{{Williams} \& {Berger}(2016)}]{willians2016}
{Williams}, P.~K.~G., \& {Berger}, E. 2016, \apjl, 821, L22

\bibitem[{{Worthey}(1994)}]{Worthey1994ApJS}
{Worthey}, G. 1994, \apjs, 95, 107

\bibitem[{Yamasaki \& Totani(2020)}]{yamasaki2020galactic}
Yamasaki, S., \& Totani, T. 2020, The Astrophysical Journal, 888, 105

\bibitem[{Yao {et~al.}(2017)Yao, Manchester, \& Wang}]{yao2017new}
Yao, J., Manchester, R., \& Wang, N. 2017, The Astrophysical Journal, 835, 29

\bibitem[{{Yaron} {et~al.}(2020){Yaron}, {Ofek}, {Gal-Yam}, \&
  {Sass}}]{2020TNSAN..70....1Y}
{Yaron}, O., {Ofek}, E., {Gal-Yam}, A., \& {Sass}, A. 2020, Transient Name
  Server AstroNote, 70, 1

\bibitem[{Yi {et~al.}(2014)Yi, Gao, \& Zhang}]{yi2014multi}
Yi, S.-X., Gao, H., \& Zhang, B. 2014, The Astrophysical Journal Letters, 792,
  L21

\bibitem[{{Zhou} {et~al.}(2020){Zhou}, {Zhou}, {Chen}, {Wang}, {Vink}, \&
  {Wang}}]{Zhou2020ApJ}
{Zhou}, P., {Zhou}, X., {Chen}, Y., {et~al.} 2020, \apj, 905, 99

\bibitem[{{Zou} {et~al.}(2019){Zou}, {Gao}, {Zhou}, \& {Kong}}]{Zou2019ApJS}
{Zou}, H., {Gao}, J., {Zhou}, X., \& {Kong}, X. 2019, \apjs, 242, 8

\end{thebibliography}

\appendix

\section{MCMC Simulation}
\label{app:mcmc}

We performed an MCMC simulation to estimate the maximum redshift of FRB 20181030A. We used a likelihood defined by the relation, DM$_{\mathrm{FRB}}$ =  DM$_{\mathrm{host}}$/(1+z) + DM$_{\mathrm{MW}}$ + DM$_{\mathrm{MW,halo}}$ + DM$_{\mathrm{IGM}}$, where DM$_{\mathrm{FRB}}$ = 103.5 $\pm$ 0.3 pc cm$^{-3}$. 
Table \ref{tab:mcmc-priors} summarizes the individual DM components and their respective priors.
Similar to \cite{Keane2016} and \cite{willians2016}, we modelled the Milky Way disk DM (DM$_{\mathrm{MW}}$) as a Gaussian with a mean equal to the minimum of the two Galactic DM model predictions = 33 pc cm$^{-3}$ (see Table \ref{tab:params}; the maximum redshift estimate would be larger, and so more conservative), and a standard deviation ($\sigma$) = 20\% of the mean DM$_{\mathrm{MW}}$ value, a commonly assumed uncertainty for both the models \citep{cordes2002ne2001,yao2017new}. Moreover, this is in agreement with the maximum DM estimate of the Milky Way disk along the FRB sight-line using the $\overline{{\rm{DM} \sin|b|}}$ estimate from \cite{ocker2020ApJ}
(see \S\ref{sec:host-search}). 
For DM$_{\mathrm{MW,halo}}$, we assumed a Gaussian distribution such that at 3$\sigma$, the DM$_{\mathrm{MW,halo}}$ is either 0 or 80 pc cm$^{-3}$. This choice is motivated to account for the large uncertainty in the Milky Way halo DM contribution \citep{keating2020exploring}. 

For DM$_{\mathrm{host}}$, we assumed a log-normal probability distribution as suggested by \cite{macquart2020census},
\begin{equation}
    \rm{p(DM_{\mathrm{host}}) = \frac{1}{DM_{\mathrm{host}} \sigma_{host} \sqrt{2\pi}}exp\Bigg[-\frac{(log(DM_{\mathrm{host}}) - \mu_{host})^{2}}{2\sigma^{2}_{host}}\Bigg]},
\end{equation}
with e$^{\mu_{\rm{host}}}$ = 68.2 pc cm$^{-3}$ and $\rm{\sigma_{host}}$ = 0.88. Similarly, for DM$_{\mathrm{IGM}}$, we use a semi-analytical model that \cite{macquart2020census} computed to quantify the uncertainty in DM$_{\rm IGM}$ at a given redshift (z):
\begin{equation}
    \rm{p_{IGM}(\Delta) = A\Delta^{-\beta}exp\Bigg[\frac{- (\Delta^{-\alpha} - C_{0})^{2}}{2\alpha^{2}\sigma^{2}_{1}}\Bigg]},
\end{equation}

\noindent where $\rm{\Delta = DM_{IGM}/\mean{DM_{IGM}(z)}}$, C$_{0}$ is the normalization constant, $\sigma_{1}$ = 0.2 z$^{-0.5}$ , $\alpha = 3$, $\beta = 3$,  and $\rm{\mean{DM_{IGM2}(z)}}$ is the average DM$\rm{_{IGM}}$ estimate which is a function of redshift and assumed cosmology\footnote{We adopt the Planck cosmological parameters \citep{2016Aplanck}.}, defined in Equation 2 of \cite{macquart2020census}. 
 
For the MCMC sampling, we used the \texttt{emcee} package \citep{fm2013}, which implements an affine-invariant sampling algorithm proposed by \cite{gw2010}. We use 256 walkers of 20,000 samples after discarding 1000 burn-in samples from each walker, and thinned the samples by a factor of 100. To assess the convergence of the samplings, we estimated the mean proposal acceptance fraction = 42\%, and the chain autocorrelation length $\approx$ 1.43. Both of the estimates are within the acceptable range. Lastly, we also estimated convergence criterion for the redshift parameter, $\hat{R} \approx 1.09$ which implies good convergence of the MCMC \citep{gelman2013}. 

From the MCMC analysis, we marginalized the redshift posterior over all other priors and calculated a one-sided 95\% Bayesian credible upper limit = 0.05. This is the maximum redshift of FRB 20181030A used in our analysis.  

\begin{table}[ht]

\caption{Parameters used in the MCMC analysis described in Appendix \ref{app:mcmc}.}
\label{tab:mcmc-priors}
\begin{tabular}{@{} *4l @{}}\toprule
\textbf{Parameter} & \textbf{Symbol} &\textbf{Units}& \textbf{Prior}\\\midrule
Host galaxy redshift & z & $-$ & U(10$^{-4}$,1)\\
Host galaxy DM & DM$_{\mathrm{host}}$& pc cm$^{-3}$ & LN(e$^{68.2}$,0.88)\\
Milky way DM & DM$_{\mathrm{MW}}$ & pc cm$^{-3}$ & N(33, 20\%$\times$33 )\\
Milky way halo DM & DM$_{\mathrm{MW,halo}}$ & pc cm$^{-3}$ & N(40,33.33\%$\times$40)\\
IGM DM & DM$_{\mathrm{IGM}}$ &  pc cm$^{-3}$ & Equation 4 from \cite{macquart2020census} \\\bottomrule 
 \hline
\end{tabular}

\end{table}

\begin{figure}[htbp]
\includegraphics[width=.95\linewidth]{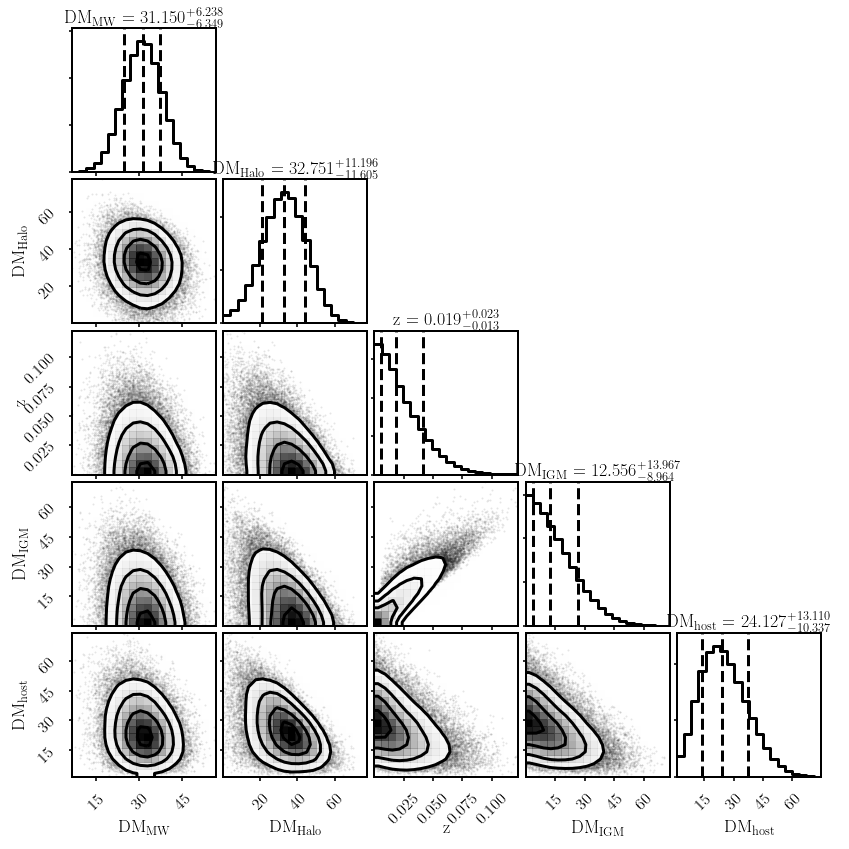}
\caption{The results of a Markov-Chain Monte Carlo (MCMC) analysis discussed in Appendix \ref{app:mcmc}. Constraints on different FRB 20181030A DM components are derived using a Bayesian framework. The marginalized distribution for each DM component is shown along the diagonal of the corner plot. All DM units are in
pc~cm$^{-3}$.} 
\label{fig:mcmc}
\end{figure} 


\section{Long-slit spectroscopy of NGC 3252}
\label{sec:longslit}

\begin{figure}[ht]%
    \centering
    \subfigure{\includegraphics[angle=-90, origin=top, width=.35\linewidth]{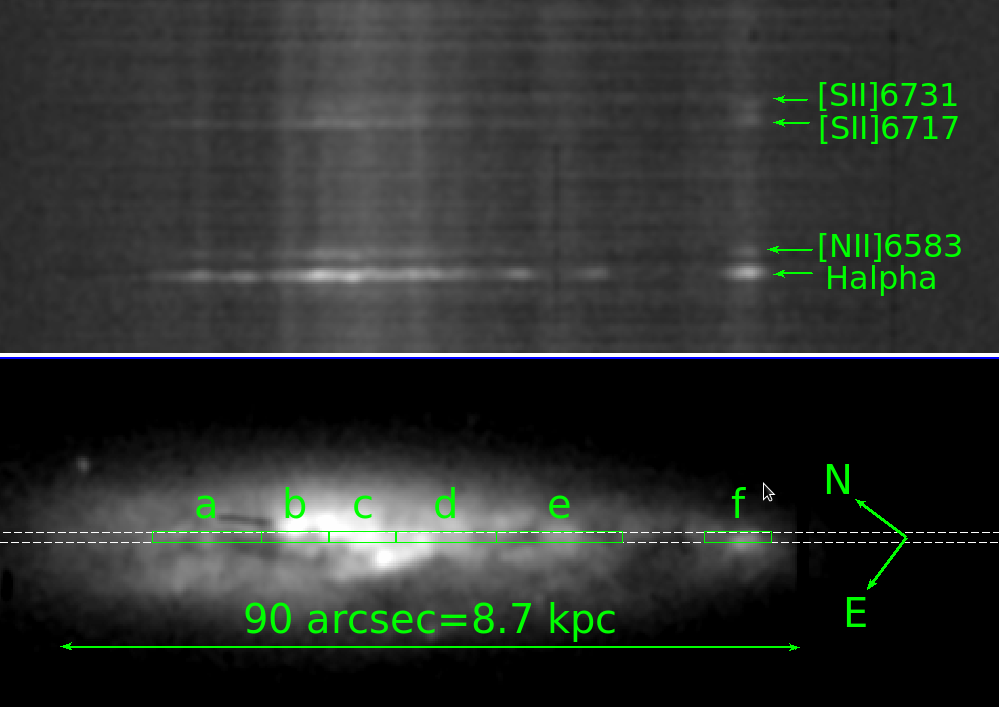}}%
    \hfill
    \subfigure{\includegraphics[width=.60\linewidth,origin=top]{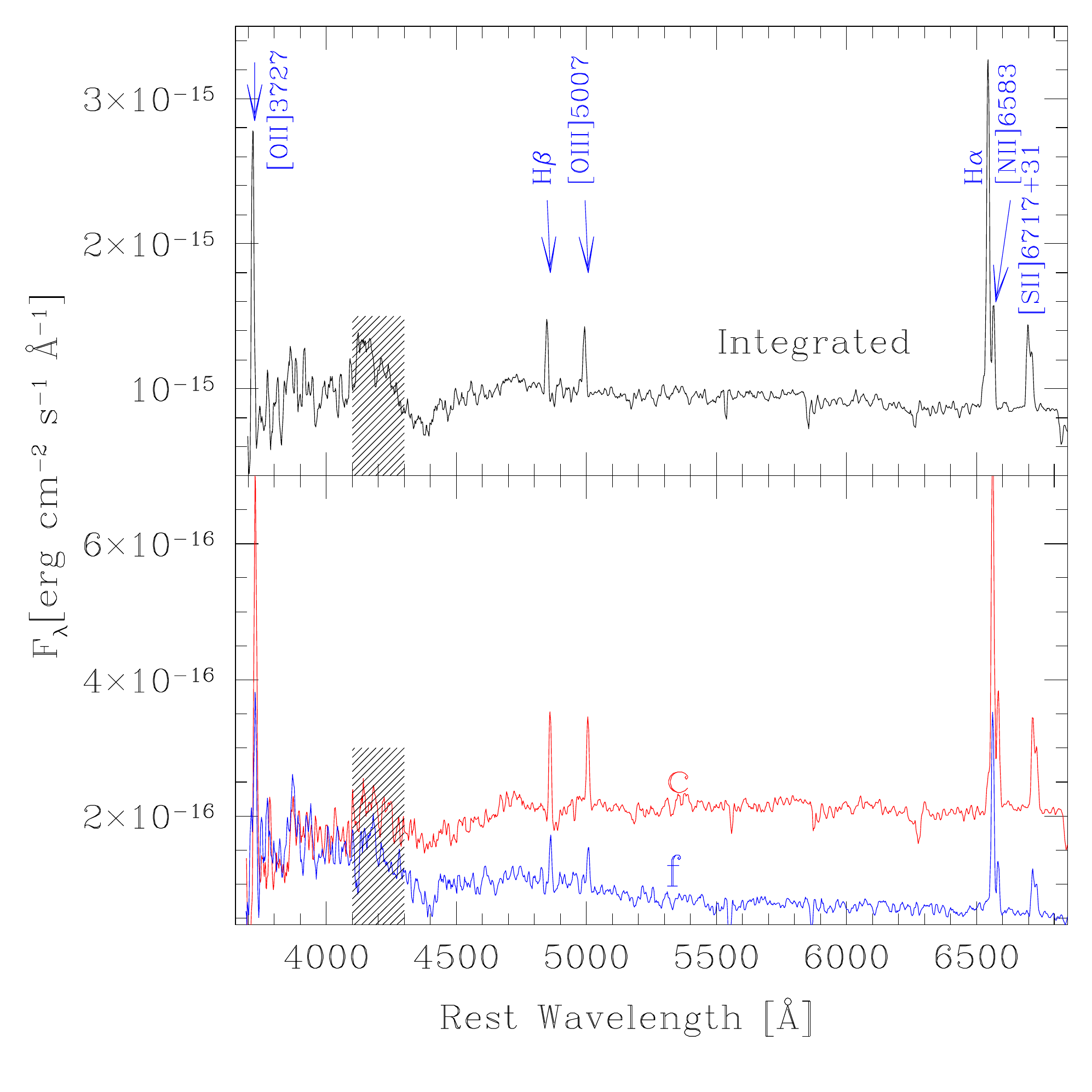}}%
    \hfill
    \caption{(Left) GTC/OSIRIS $r$-band acquisition image showing the position of the long-slit (white dashed rectangle; PA=37.31$^\circ$) and the zones a--f selected for the spectral extraction (green rectangles). The image scale and orientation are shown. (Center) Spectral image along the long-slit 
    showing prominent emission lines in the red part of the spectrum.
    (Right) Rest frame spectra, where the top spectrum was obtained by summing the rest frame spectra of all the six zones denoted a--f. Spectra marked ``c" and ``f" correspond to the zone passing through the nucleus and a bright H{\sc ii}, respectively. 
    Note that the latter region lies almost at the center of the 90\% localization ellipse of the FRB source shown (see Figure~\ref{fig:frbfov}).
    A bump in the spectra between 4100~\AA\ and 4300~\AA\ is due to a detector artefact, which is shown by shaded box.}
    \label{fig:longslit-ngc3252}%
\end{figure}


We acquired the long-slit data of NGC 3252 in order to estimate the physical properties of NGC 3252, such as nebular metallicity and dust extinction. Here we discuss the steps for reducing the NGC 3252 longslit spectroscopy data. In Figure~\ref{fig:longslit-ngc3252}, we show the zones used for the extraction of spectra along the long-slit. H$\alpha$ and other nebular lines are traced over a zone of $\sim$90\arcsec\ (8.7~kpc) which covers the entire bright optical extent of the galaxy. Several emission knots are seen along the slit, especially in the H$\alpha$ spectral image. Each knot represents an H{\sc ii} region in the host galaxy. The presence of these H{\sc ii} regions allows us to obtain physical quantities along the slit using the physics of photoionized nebulae \citep{Osterbrock2006}. In order to maximize the S/N ratio of the extracted spectra, we defined six zones, identified by letters a--f, in such a way that each zone contains at least one of the emission knots. The zone c spectrum contains the nucleus, and the zone f spectrum corresponds to a bright H{\sc ii} region to the south-west of the nucleus. This H{\sc ii} region lies almost at the center of the 90\% localization ellipse of the FRB source shown in Figure~\ref{fig:frbfov}. In the right panel, we show the extracted spectrum for these two regions. Each extracted spectrum was analyzed to measure the fluxes of bright nebular lines using the Gaussian fitting technique of the {\it splot} task in {\sc iraf}, which also performs a measurement and subtraction of continuum flux. The prominent lines in the spectrum were identified and are shown in the top panel. A deblending algorithm was used to extract accurate fluxes of [NII]$\lambda$6548 and [NII]$\lambda$6583 lines in the presence of the bright H$\alpha$ line and also to resolve the [SII] doublet. 

The measured line fluxes in each spectrum are given in Table~\ref{abundance}. The spectra were first corrected for Doppler shift using a mean of the recessional velocities measured using the H$\beta$, H$\alpha$ and [OIII]$\lambda$5007 lines in each spectrum. The measured velocities are also given in Table \ref{abundance}. All the six rest frame spectra were summed to get an integrated spectrum of the galaxy, which is shown in the top-right panel in Figure~\ref{fig:longslit-ngc3252}. Values measured for the integrated spectrum are given in the last column of Table \ref{abundance}. The mean of the velocities of
the six extracted spectra was taken as the velocity of the integrated spectrum, which agrees very well with the systemic velocity of 1156$\pm$6~km\,s$^{-1}$ reported in NED \citep{1992ApJS...81....5S}.
The H$\alpha$ and H$\beta$ emission line fluxes stated in Table \ref{abundance} are used to obtain the visual extinction $A_V$ experienced in each zone, shown in Figure \ref{fig:longslit-ngc3252}, following the Balmer decrement method for case B recombination of a typical photo-ionized nebula \citep[$T_{\rm e}$ =10000~K, $n_{\rm e}$=100 cm$^{-3}$;][]{Osterbrock2006} and the reddening curve of \cite{Cardelli1989}. We corrected the observed H$\alpha$ and H$\beta$ fluxes for the effects of the underlying stellar absorption by assuming an absorption equivalent width of 2~\AA\ following \cite{mccall1985ApJS}.
The resulting $A_V$ values vary between  0.8--1.6~mag in the zones along the slit. The line fluxes were corrected for the measured extinction and are given as ratios with respect to the flux of the H$\beta$ line, which is multiplied by 100 following the normal convention.

Electron temperature-sensitive Auroral lines were not detected in any of the extracted spectra. However, nebular lines for the determination of the oxygen and nitrogen abundances using the strong-line method
are detected with S/N $>$10. We used the calibrations of \citet{Pilyugin2016} 
for this purpose (their Equations 4 and 13). The resulting values of 12+$\log$(O/H) 
are given in Table \ref{abundance}
for each zone as well as that measured in the integrated spectrum. 

\begin{table}[ht]
\caption{Physical quantities from long-slit spectra of NGC 3252.}
\label{abundance}
\begin{center}
\resizebox{1.00\textwidth}{!}{ 
\hspace{-1.0in}
\begin{tabular}{lccccccc} \hline
\toprule
\textbf{Quantity}$^a$ & \textbf{a} & \textbf{b} & \textbf{c} & \textbf{d} & \textbf{e} & \textbf{f} & \textbf{Integrated} \\
\midrule
   R.A.(J2000)         &   10:34:25.50         &10:34:23.94            &   10:34:22.73        & 10:34:21.27           &10:34:19.26            & 10:34:16.11         &                    \\
   Dec.(J2000)       &   +73:46:05.0         &+73:45:56.4            &   +73:45:49.7        & +73:45:41.6           &+73:45:30.5            & +73:45:13.1         &                    \\
   Area[$^{\prime\prime}\times^{\prime\prime}$]  &    13.3$\times$1.2  & 8.2$\times$1.2   &   8.2$\times$1.2   &  12.3$\times$1.2  & 15.3$\times$1.2    & 8.2$\times$1.2    & 65.5$\times$1.2      \\
   \hline
   I([OII]3727)  &  38.5  $\pm$  4.6  &  556.7 $\pm$   3.7  &  498.2 $\pm$  11.1 &  769.8 $\pm$   9.5 &   412.4 $\pm$   8.2 &  168.7  $\pm$  0.3 &   437.0 $\pm$  4.5  \\
  I(H$\beta$)    &   100.0            &   100.0             &  100.0             &  100.0             &   100.0             &  100.0               & 100.0          \\
  {}I([OIII]5007)&  79.8  $\pm$  0.1  &   51.7 $\pm$   4.0  &   84.2 $\pm$   2.8 &   89.3 $\pm$   1.3 &    51.4 $\pm$   5.6 &   74.0  $\pm$  0.5 &    77.4 $\pm$  1.0  \\
   I(H$\alpha$)  & 287.0  $\pm$ 53.1  &  287.0 $\pm$  66.3  &  287.0 $\pm$  29.8 &  287.0 $\pm$  38.8 &   287.0 $\pm$  75.0 &  287.0  $\pm$ 37.4 &   287.0 $\pm$ 32.2  \\
   {}I([NII]6583)&  96.1  $\pm$ 14.1  &   82.7 $\pm$  16.1  &   89.9 $\pm$   7.6 &   87.7 $\pm$   9.3 &    69.0 $\pm$   8.2 &   61.9  $\pm$  4.1 &    80.6 $\pm$  7.0  \\
   {}I([SII]6717)&  53.8  $\pm$  5.9  &   61.7 $\pm$   9.5  &   62.2 $\pm$   4.1 &   67.8 $\pm$   6.7 &    58.2 $\pm$   2.9 &   63.5  $\pm$  3.0 &    60.6 $\pm$  3.7  \\
   {}I([SII]6731)&  25.1  $\pm$  0.1  &   42.5 $\pm$   4.6  &   47.4 $\pm$   2.1 &   45.9 $\pm$   3.2 &    44.9 $\pm$   1.2 &   41.6  $\pm$  0.1 &    45.0 $\pm$  0.3  \\
\hline
logF(H$\beta_0$)[erg\,cm$^{-2}$\,s$^{-1}$] 
                 &  $-$14.50$\pm$0.20 &  $-$14.25$\pm$ 0.19 & $-$14.25$\pm$ 0.11 &$-$14.02$\pm$0.14   & $-$14.28$\pm$   0.29& $-$14.62$\pm$ 0.16 & $-$13.50 $\pm$0.11   \\
   $A_V$[mag]    &     0.9  $\pm$0.4  &    1.4  $\pm$  0.4  &   0.9   $\pm$ 0.2  &    1.6 $\pm$ 0.2   &     1.6 $\pm$  0.4  &  0.8 $\pm$    0.3  &  1.3    $\pm$ 0.2   \\
EW(H$\beta$)[\AA]&    12.4            &    11.1             &     7.3            &    3.3             &     2.9             &  8.6               &    5.8                 \\
Velocity [km\,s$^{-1}]$ 
                &  1073 $\pm$43      &  1105 $\pm$49       & 1056 $\pm$  35     & 1125 $\pm$  15     & 1166 $\pm$ 57       & 1274$\pm$ 50       &   1133 $\pm$  79 \\
12+log(O/H)      &   8.64 $\pm$ 0.05  &     8.38$\pm$ 0.05  &    8.46  $\pm$0.05 &    8.38 $\pm$0.05  &    8.37  $\pm$0.05  &  8.49   $\pm$0.05  &   8.44   $\pm$  0.05   \\
\hline
\end{tabular}
}

\end{center}

$^a$ In the first block, center coordinates of the rectangular zones (named a to f; see Figure~\ref{fig:fig}) chosen for the extraction of spectra are given. The next block contains the extinction corrected 
fluxes of prominent nebular lines relative to the flux of the H$\beta$ line, i.e. I($\lambda$)=100$\times$ F($\lambda$)/F(H$\beta$).
The last block contains the extinction-corrected H$\beta$ flux and physical quantities derived from the diagnostics of the nebular lines.
\end{table}


\section{Stellar Population Synthesis Using {\tt Prospector}}
\label{app:prospector}

We infer several physical properties of NGC 3252 (at 20 Mpc) using a python-based Bayesian inference code, {\tt Prospector} \citep{Leja2017,prospect2019}, which estimates galaxy properties using stellar population synthesis models defined within the framework of the Flexible Stellar Populations Synthesis (FSPS) stellar populations code \citep{Conroy2009}. {\tt Prospector} provides an MCMC framework via \texttt{emcee} to fit observed spectral energy distributions (SEDs) and estimate posterior distribution for each free-parameters. In this paper, we use {\tt Prospector} to estimate the stellar mass, metallicity, and mass-weighted stellar population age of NGC 3252. We use 17 broadband filters from the GALEX FUV filter at 1549 \AA\ through the Herschel telescope bands that provide coverage at far-infrared wavelengths
as shown in Figure \ref{fig:sed} and Table \ref{sed:maggies}.

\begin{table}[htbp]

\caption{ Free parameters and their associated priors for the {\tt Prospector} `delayed-$\tau$' model.}
\label{tab:sfhmodel}
\begin{center}
\hspace{-1.in}
\begin{tabular}{ p{2cm}p{6.5cm}p{5.5cm}}
 \hline
 Parameter & Description & Prior\\
 \hline
 log(M/M$_{\odot}$)   &  total stellar mass formed & uniform: min=8, max=11\\
 log(Z/Z$_{\odot}$)  &   stellar metallicity   & Gaussian: mean=-0.25, $\sigma$=0.21 \\
 dust2 & diffuse V-band dust optical depth  &  top-hat: min=0.0, max=3.0\\
 t$_{\rm age} \rm{~[Gyrs]}$   & stellar population age of NGC 3252 &top-hat: min=0.01, max=13.6\\
 $\tau \rm{~[Gyrs]}$ & e-folding time of the SFH   & uniform: min=0.1, max=30\\
 \hline
\end{tabular}
\end{center}
\end{table}

\begin{table}[ht]
\begin{center}
\hspace{-1.in}
\caption{17 broadband filters used to model the SED of NGC 3252.}
\label{sed:maggies}
\begin{tabular}{@{} *4l @{}}
\toprule
Instrument$^c$ & Filter & Effective Wavelength & Flux density$^{a,b}$\\
& & \AA & maggies \\ \midrule 

GALEX & FUV & 1549 & 4.38 $\times 10^{-7}$ \\
 & NUV & 2304 & 5.56 $\times 10^{-7}$  \\
DESI$^{d}$ & g & 4670 & 4.98 $\times 10^{-6}$  \\
 & r & 6156 & 9.26 $\times 10^{-6}$ \\
 & z & 8917 & 1.41 $\times 10^{-5}$  \\
2MASS & J & 12319 & 1.75 $\times 10^{-5}$   \\
 & H & 16420 & 1.79 $\times 10^{-5}$   \\
 & Ks & 21567 & 1.67 $\times 10^{-5}$   \\
WISE & W1 & 33461 & 9.75 $\times 10^{-6}$   \\
 & W2 & 45952 & 6.33 $\times 10^{-6}$  \\
 & W3 & 115526 & 1.99 $\times 10^{-5}$   \\
 & W4 & 220783 & 2.50 $\times 10^{-5}$   \\
Herschel & PACS(Green) & 979036 & 8.40 $\times 10^{-4}$   \\
 & PACS(RED) & 1539451 & 1.13 $\times 10^{-3}$   \\
 & SPIRE(PSW) & 2428393 & 5.87 $\times 10^{-4}$   \\
 & SPIRE(PMW) & 3408992 & 3.00 $\times 10^{-4}$   \\
 & SPIRE(PLW) & 4822635 & 1.27 $\times 10^{-4}$   \\\bottomrule 
 \hline
\end{tabular}
\end{center}

$^{a}$  Note that 1 maggie is defined as the flux density in Janskys divided by 3631. Fluxes at $ \lambda < 100000$ \AA~are corrected for Galactic extinction according to the prescription of \cite{schlafly2011ApJ}.\\
$^b$ All broadband fluxes are assigned a 20\% fractional uncertainty.\\
$^c$ Except for the DESI survey, all instruments' flux densities are obtained from the aperture-matched photometry catalog of nearby galaxies by \cite{clarke2018A&A}.
\\$^{d}$ For DESI filter magnitudes, we used the photometric redshift catalog by \cite{Zhou2020ApJ}.
\end{table}

All flux densities are estimated after correcting for the Milky Way extinction. We fit a delayed-$\tau$ star formation history model \citep{simha2014,carnall2019ApJ} with five free parameters $-$ metallicity, mass-weighted stellar population age, star formation timescale, and V-band optical extinction (= 1.086 $\times$ dust2), which are described in Table~\ref{tab:sfhmodel}.
In this model, the star-formation history is proportional to t$\rm{\times exp(-t /\tau}$), where t is the time since the formation epoch of the galaxy, and $\tau$ is the characteristic decay time of our star-formation history. Additionally, we enabled nebular emission \citep{byler2017} and dust emission \citep{draine2007} models in the FSPS framework along with a standard dust attenuation model from \cite{Calzetti2000ApJ}.
Finally, all five parameters are given standard {\tt Prospector} priors, except log(Z/Z$_{\odot}$), which is informed by the constraints derived in \S\ref{sec:physical-properties}. This is included to reduce the effect of the age-metallicity degeneracy \citep{Worthey1994ApJS}.  
We used a Gaussian prior to model log(Z/Z$_{\odot}$) with mean $= -0.25$ derived in \S\ref{sec:physical-properties} using optical spectral lines. However, we increased the $\sigma$ value by a factor of three to account for any potential bias in converting the oxygen abundance to nebular metallicity, a conservative choice given that the conversion error is typically $\sim 0.02$ \citep{2009ApJ...705L.123S}, which is less than one sigma error on the $\mathrm{\log(Z_{\rm gas}/Z_\odot)}$, i.e. 0.07 (see Table \ref{tab:host-properties}).

Using this framework, we derived a metallicity fraction, $\mathrm{log}(\text{Z}/\text{Z}_{\odot})$ =  $-0.21^{+0.18}_{-0.19}$, and the present-day stellar mass of the galaxy = $ 5.8^{+1.6}_{-2.0} \times 10^{9} \rm{M}_\odot$. To estimate the best-fitted mass-weighted stellar population age value, we used Equation 5 of \cite{carnall2019ApJ} and found it to be 4.8$^{+1.6}_{-1.8}$ Gyr.
All these values are provided in Table \ref{tab:host-properties}. Note that the quoted uncertainties in all cases are 1$\sigma$ values.

\normalsize
\end{document}